 \documentclass[12pt,preprint,onecolumn]{emulateapj}
\usepackage{enumerate}

\usepackage{amsmath}
\usepackage{amssymb}

\usepackage{lineno}

\usepackage{rotating}

\usepackage{soul}
\usepackage[usenames,dvipsnames]{color}

\usepackage{url}

\usepackage{hyperref}
\hypersetup{
colorlinks=true,
breaklinks=true,
urlcolor= RedViolet,
citecolor=blue,
linkcolor=blue,
breaklinks=true
}
\usepackage{breakurl}

\usepackage{setspace} % interligne
\usepackage[framemethod=tikz]{mdframed}

\definecolor{vertdc1}{RGB}{20,89,33}
\definecolor{blood}{RGB}{193,41,41}
\definecolor{viol}{RGB}{109,10,186}
\definecolor{dgreen}{RGB}{9,95,29}
\definecolor{dorange}{RGB}{197,69,6}

\definecolor{CNRSBlue}{RGB}{26,48,81}
\definecolor{CNRSLightBlue}{RGB}{10,141,167}
\definecolor{DarkRed}{RGB}{212,0,0}
\definecolor{DarkRed2}{RGB}{150,0,0}
\definecolor{Violet}{RGB}{122,47,214}

\newcommand{\GG}[1]{} % pour intervertir les citations de Malaska et al (2017) (ne pas avoir le Press Release avant le papier)
\slugcomment{}

\newcommand{\trait}{\begin{center}\hrule\end{center}}

\shorttitle{}

\shortauthors{Coutelier et \textit{al.}}

\begin{document}
%\linenumbers


\title{Distribution and intensity of water ice signature in South Xanadu and Tui Regio}

\author{
Ma\'{e}lie~Coutelier\altaffilmark{1},
Daniel~Cordier\altaffilmark{1},
Beno\^{i}t~Seignovert\altaffilmark{2},
Pascal~Rannou\altaffilmark{1},
Alice~Le~Gall\altaffilmark{3}$^,$\altaffilmark{4},
Thibaud~Cours\altaffilmark{1},
Luca~Maltagliati\altaffilmark{5},
S\'{e}bastien~Rodriguez\altaffilmark{6},
}

\altaffiltext{1}{Groupe de Spectrom\'{e}trie Mol\'{e}culaire et Atmosph\'{e}rique - UMR CNRS 7331
               Campus Moulin de la Housse - BP 1039
               Universit\'{e} de Reims Champagne-Ardenne
               51687 REIMS -- France}

\altaffiltext{2}{Université de Nantes, Univ Angers, CNRS, LPG, UMR 6112, F-44000 Nantes, France}

\altaffiltext{5}{Nature Publishing Group, LONDON, UNITED-KINGDOM}
\altaffiltext{3}{LATMOS/IPSL, UVSQ Universit\'{e} Paris-Saclay, Sorbonne Universit\'{e}, CNRS, Paris, France}
\altaffiltext{6}{Universit\'{e} de Paris, Institut de physique du globe de Paris, CNRS, F-75005 Paris, France}
\altaffiltext{4}{Institut Universitaire de France (IUF), Paris, France}    

\email{maelie.coutelier@univ-reims.fr}
                 
%\begin{linenumbers}                 
\begin{abstract}
  %{\noindent\textbf{Context} --- .}\\
  %{\noindent\textbf{Aims} --- . }\\
  %{\noindent\textbf{Methods} --- .}\\
 % {\noindent\textbf{Results} --- .}\\

  Titan's surface was revealed by \textit{Cassini}'s instruments, showing the presence of liquid hydrocarbons in lakes, and features like dry riverbed. In order to study the sediment transport in Titan's channels and to map distribution of the water-ice signature in these terrains, we use a radiative transfer model to retrieve the surface albedo, after we estimated VIMS error \textit{via} an original method. We also establish a criteria related to the intensity of the water ice signature. The tuning of the radiative transfer model shows that the fractal dimension of Titan's aerosols is higher than previously thought, around 2.3 - 2.4. We find spots of increasing signal of water ice downstream, at the margins of Tui Regio, that could correspond to alluvial fans, deltas or crater rims. We also observe a very low water ice signal on Tui Regio, with a positive gradient between the central region and the boundary of the area, possibly due to the thickness variation of an evaporitic layer. The riverbeds show within the error bars a decreasing grain size from the top to the bottom of the channels.

\end{abstract}

\newpage

%%%%%%%%%%%%%%%%%%%%%%%%%%%%%%%%%%%%%%%%%%%%%%%%%%%%%%%%%%%%%%%%%%%%%%%%%%%%%%%%%%%%%%%%%%%%%%%%%%%%%%%%%%%%%%%%%%%%%%%%%%%%%%%%%%%%
\sffamily
\section{Introduction}\label{intro}

% Quelques généralités sur Titan :
Titan, the biggest moon of Saturn is one of the most complex moon of the Solar System, with dense atmosphere, hydrocarbon cycle, and water-ice crust. Titan has been targeted by two major space missions: \textit{Voyager I} (1980) and \textit{Cassini-Huygens} (2004-2017), and  is still the subject of many studies. Lakes and seas of liquid hydrocarbons were discovered 
by \textit{Cassini}'s ISS and RADAR in Titan's polar regions \citep{stofan_etal_2007,turtle_etal_2009}. The RADAR instrument also revealed in equatorial regions geomorphological structures related to the presence of liquid, like fluvial valleys incised in the \emph{bedrock}, and alluvial fans \citep{legall_etal_2010,Birch2016}. The existence of evaporitic terrains where also suggested \citep{barnes_etal_2011,cordier_etal_2013b,macKenzie_etal_2014}, often in place of paleo-sea \citep{Moore2010}. Water ice signature is not ubiquitous on Titan, contrary to most of Saturn's and Jupiter's moons, like Encelade and Europa. It was detected with \textit{Cassini}'s Visual and Infrared Mapping Spectrometer (VIMS) in Titan's infrared dark region, often at the transition between dark and bright unit, and mixed with a dark material \citep{mccord_etal_2006,Soderblom2007,brossier2018}. \citet{griffith_etal_2019} also highlighted an equatorial corridor of possibly exposed water-ice using a principal component analysis (PCA) technique applied on VIMS data, showing a large scale terrains with low and high water ice content. While PCA is extremely useful for large data sets, and to study weak feature, it is not as detailed as a radiative transfer modeling and can not be used to retrieve quantitative information. 

\citet{barnes_etal_2007} have studied the channels of Mabon Macula using a combination of Synthethic Aperture Radar Image (SAR) and VIMS measurements. They find that the spectral signals of the channels share characteristics with Titan's \textquoteleft\textquoteleft dark blue\textquoteright\textquoteright~ terrains according to the adopted color composition. The considered terrains show an enrichment in water ice \citep{brossier2018,rodriguez2006} and the presence of benzene. These dark blue areas should result from the transport of material eroded from the mountains by the channels.This has been confirmed by \citet{langhans_etal_2012} and \citet{brossier2018}. The latter studied the \textquoteleft\textquoteleft dark blue\textquoteright\textquoteright~ terrains found with VIMS on Titan. They used an accelerated version of radiative transfer model with precalculated tables to compare the spectral variations of different terrains. The IR-blue units were found to be enriched in water-ice (20 to 50 \%), even if it is not the main component. \citet{brossier2018} also described different types of IR-blue units. First, they have identified outwashed plains that are located between IR-bright terrains and dune field, created by sediments transported from mountain IR-bright units. Secondly, they found patches of IR-blue units located inside IR-bright terrains, produced by the erosion of these mountainous terrains. The last type is the ejecta blanket from impact craters. They also found that rivers are IR-blue, probably enriched in water-ice pebbles as well.

We focused our study on Titan's river bed, where the erosion due to liquid hydrocarbon could have revealed the bedrock, and where we could have clues on the sediment transport. Here we propose a study of the water ice signal distribution in South Xanadu and Tui Regio, where there are river channels \citep{legall_etal_2010}, paleo-seas of hydrocarbons \citep{hofgartner_etal_2020}, and terrains classified as IR-blue patches by \citet{brossier2018}. We employ a photometric analysis combined with an original method based on indicators to characterize the relative strength of the water-ice signature on the chosen VIMS cubes.

We focused on two radar bright channels investigated by \citet{legall_etal_2010}, larger than those reported by \citep{barnes_etal_2007} in Mabon Macula. These channels are located south of Xanadu, emerging directly from a mountain range. They are at the north of the evaporitic region Tui Regio \citep{MacKenzie2016}. They form dry dendritic fluvial valleys \citep{langhans_etal_2012}, with a width up to $8$~km. Some areas studied in this work have also been investigated by \citet{griffith_etal_2019}, where evaporitic terrains were classified as ice-poor.
%
%
%===================================================================================================================================
%\begin{figure}[!t]
%\begin{center}
%\includegraphics[angle=0, width=8 cm]{fig/binary_N2-CH4_5.ps}
%\caption[]{\label{binaryN2CH4}Comparison between experimental data for the binary system N$_2$-CH$_4$ and our PC-SAFT 
%           based model \citep{cordier_etal_2017a}, 
%           for two temperatures: \REVI{$91$ K (circles)} and $95$ K (triangles). Laboratory measurements, already used by \citet{tan_etal_2013}, \REVI{come from 
%           various sources: \citet{sprow_prausnitz_1966} for $91$ K,} and \citet{parrish_hiza_1974} for $95$ K (triangles). 
%           Squares represent N$_2$ dissolution data from recent work \citep{malaska_etal_2017} respectively at $89\pm 0.5$ K and $95\pm 0.5$ K.
%           The pressure $P= 1.5$ bar represents the value determined by {\it Huygens} at ground level.}
%\end{center}
%\end{figure}
%===================================================================================================================================
% 
%%%%%%%%%%%%%%%%%%%%%%%%%%%%%%%%%%%%%%%%%%%%%%%%%%%%%%%%%%%%%%%%%%%%%%%%%%%%%%%%%%%%%%%%%%%%%%%%%%%%%%%%%%%%%%%%%%%%%%%%%%%%%%%%%%%%
%% Description des données :
\section{Radar and Infrared Observations}

To investigate our region of interest, we used two complementary instruments of the \textit{Cassini}'s mission: the RADAR, and VIMS.
\textit{Cassini}'s RADAR had four different modes: it could operate as a SAR, scatterometer, altimeter and radiometer at a wavelength of 2.2 cm. In SAR mode, it provided images of the surface that are sensitive to local slope and terrains roughness: smooth surfaces appear dark while rough terrains are radar-bright. About 70\% of Titan's surface has been mapped with this mode with a resolution ranging between 300 m/pixel and 4 km/pixel, revealing channels, lakes, mountains and dune fields.

VIMS Infra-Red instrument onboard \textit{Cassini} scanned in a spectral range from 0.88 $\mu$m to 5.12 $\mu$m. %The Visible and Infra-red Mapping Spectrometer (VIMS) instrument onboard \textit{Cassini} spacecraft takes pictures in a spectral range from 0.315 $\mu$m to 5.12 $\mu$m.
In Titan's case, due to the methane absorption, VIMS could only observe the surface in $7$ atmospheric windows centered at  $0.93$, $1.08$, $1.27$, $1.59$, $2.01$, $2.78$ and $5$~$\mu$m \citep{sotin_etal_2005,barnes_etal_2005}. Parts of the spectrum not located in an atmospheric window are called \emph{bands}.% However this is also a general term designating VIMS spectral channels. In our study, we limited our effort to the infrared part of the spectrum where the windows are located.
As the RADAR spatial resolution was higher than VIMS, the SAR map allows a better characterization of geomorphologic structures in VIMS cubes.
\begin{figure}[h]
	\begin{center}
		\includegraphics[width=16 cm, angle=0]{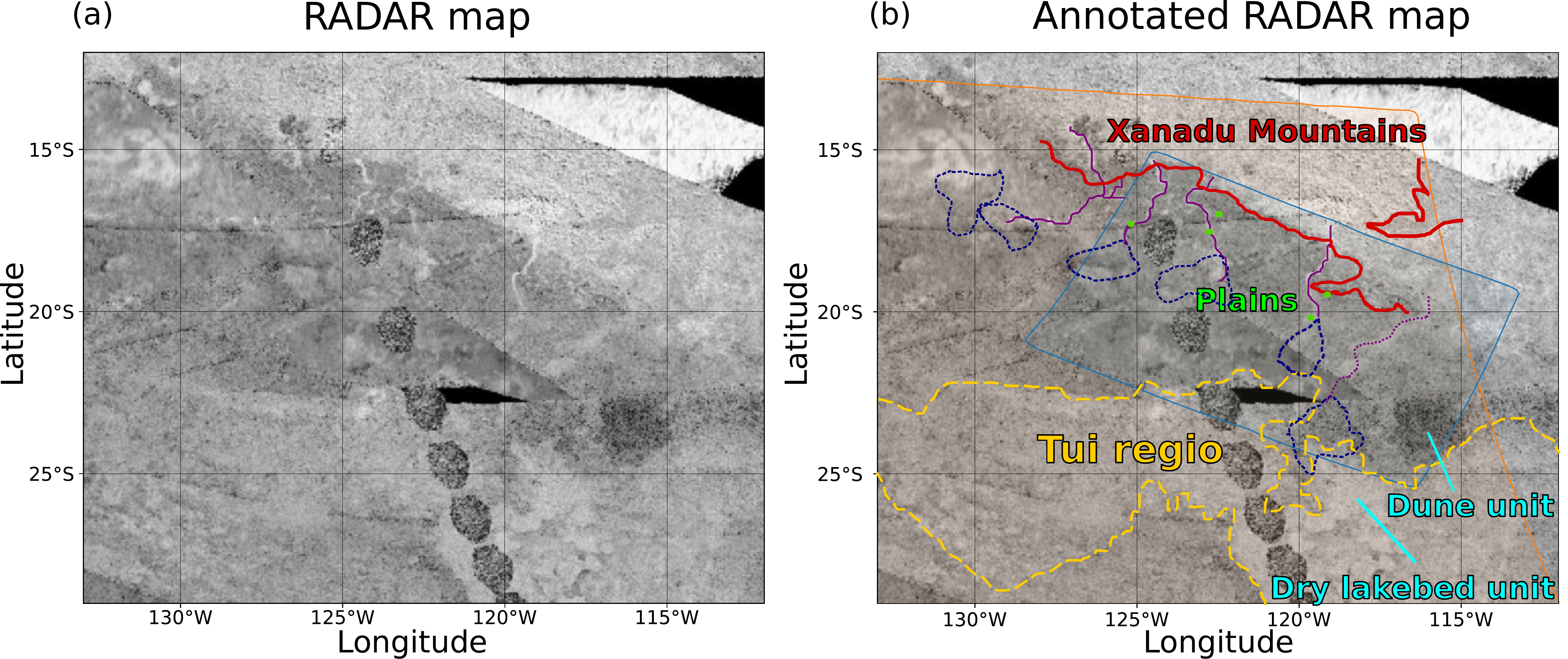}
	\end{center}
	\caption[]{\label{radarVIMS} Grey-scale: (a) RADAR swath containing the T48 and T44 flyby \citep{legall_etal_2010}. (b) Annotated RADAR map, with footprints of VIMS cube \texttt{1590648776\_1} (C15, blue) and \texttt{1809727868\_1} (C18, orange). Purple lines: the rivers bright channels on which we focus our study. Tui Regio is roughly delimited in yellow, Xanadu mountains are similarly delimited in red, and the units defined by \citet{Lopes2019} and \citet{Moore2010} are annotated in cyan. Green dots: channels that end in fan \citep{radebaugh2018}. Blues lines: areas of high water-ice signal. The row of dark spots cutting the map vertically in half comes from HiSAR data, where the surface resolution is lower than SAR data. As an other useful central portion of this map comes from HiSAR data, we have chosen to keep the complete map instead of working only with SAR flyby.  }
\end{figure}

The radar-bright river channels from \citet{legall_etal_2010} have been imaged during the T44 flyby (2008-28-05) using SAR mode. We employed the T44 SAR image to select VIMS images overlapping these channels. Among them, we chose the cubes with the highest resolution in this area.
Only two cubes have a pixel scale smaller than 18 km/pixel (see Tab. \ref{cubes} and Fig.~\ref{radarVIMS}), which is twice the width of the bright channels. A weaker resolution may not contain enough channel signal to be distinguished. The first cube 1590648776\_1 (labeled C15 hereafter) was acquired in 2008 during the same flyby as our reference radar swath. For this cube, VIMS was configured in High Resolution mode which 
%decreases the pixel binning in the sample direction to 
increases the spatial resolution in the samples direction. The second cube (C18 hereafter) was taken in normal mode 7 years later in 2015 during the T111 flyby. For C18 we do not work with the entire cube, only a square made of 38 rows of samples and lines located in our region of interest. These cubes were calibrated with the \texttt{RC19} calibration published by  \citet{clark_etal_2018}\footnote{\url{https://pds-imaging.jpl.nasa.gov/data/cassini/cassini_orbiter/vims-calibration-files/}}, using ISIS3\footnote{\url{https://isis.astrogeology.usgs.gov/}} USGS tools. They are available at the {\it Cassini VIMS Data Portal}\footnote{\url{https://vims.univ-nantes.fr}} \citep{lemouelic_etal_2019}.
% ----------------------------------------------------------------------------------
%\subsection{Georeferencing}
The navigation of the VIMS-IR cubes were computed using the latest C-smithed kernels (December 2017) and the python SpiceyPy toolkit \citep{Annex2020}. We took into account the offseted pointing boresight and the corrected initial scan mirror sample position for cubes acquired in HighRes mode, as recently reported by \citet{Nicholson2020}. Where changes exist, they are about 1 normal VIMS pixel compared to the original ISIS navigation and within the expected accuracy of the Cassini kernels ($\sim$0.12 mrad).
\begin{table}[h]
	\begin{center}		\caption[]{\label{cubes}VIMS cubes details. C15 And C18 are used in the study of the bright channels, while C14 is employed to test the validity of the model at \textit{Huygens} landing site.}
		\vspace*{0.5cm}
		\begin{tabular}{|l|l|l|l|l|l|l|l|}
			\hline
			Cube number & Name in &   Flyby   &  Date  & Exposure  & Mean ground &Size& Purpose\\
			& the text & & (DD-MM-YYYY) &  (ms)  &  resolution & (Samples x Lines) &\\
			\hline
			1590648776\_1	&	C15 &  T44 &  28-05-2008   & 325    &  15  km/pxl &	(28 x 14)&Bright channels\\
			1809727868\_1 & C18 & T111  &   07-05-2015  &  122   &    14  km/pxl  & (64 x 64)&	Bright channels\\
			1481624349\_1 & C14 & TB &  13-12-2004  & 160    &     16 km/pxl  & (24 x 12)	& Model test\\
			
			\hline
		\end{tabular}
		
	\end{center}

\end{table}
% ----------------------------------------------------------------------------------
%
%===================================================================================================================================

\begin{figure}[h]
	\begin{center}
		\includegraphics[width=16cm]{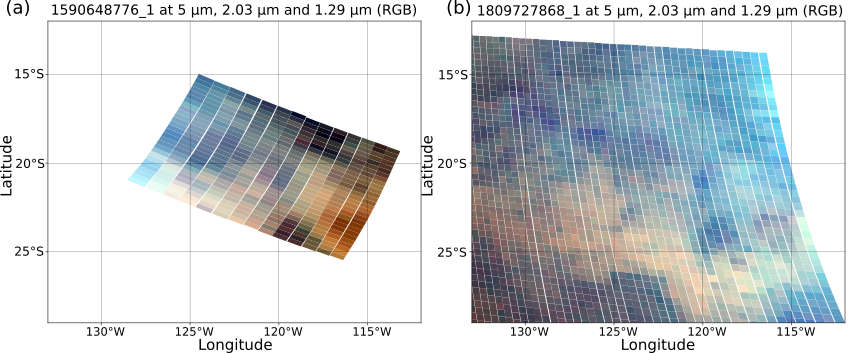}
	\end{center}
	\caption[]{\label{Barnes} Equirectangular projection of (a) C15 and (b) C18 using \citet{Barnes2007b} colors: 5.0 $\mu$m as red, 2.00 $\mu$m as green, and 1.29 $\mu$m as blue. This composition highlights evaporites in bright peach, sand in dark brown, enhanced water ice signal in dark and light blue. }
\end{figure}

C15 and C18 are re-projected in a equirectangular view for better viewing and comparison purpose in Fig.~\ref{Barnes}, with a color composition similar to that of \citet{Barnes2007b}.
%==================================================

%Two methods where used to georeference the VIMS cubes. One is \textit{via} the program ISIS, the other is a customized method. A problem arised for the cube 1590648776\_1, where the sampling mode was in high resolution. We have a 2-samples shift between the 2 methods, shifting the pixels' positions along the rows of samples. Unfortunately, we were not able to determine with absolute certainty which method was right. However, based on the superposition of the cubes in a equirectangular projection, with the customized method we to have a perfect continuity of the structures. Therefore, we think that the customized method is right, and that ISIS may have a problem with georeferencing in high-resolution sampling mode.

% 
%%%%%%%%%%%%%%%%%%%%%%%%%%%%%%%%%%%%%%%%%%%
%% Description du modèle :
\section{The Radiative Transfer Model}

In order to retrieve and study the surface albedo in different areas of interest from VIMS-IR spectra, we used a model of photometry inherited from \citet{rannou_etal_2010,rannou_etal_2016}. We first defined the average scattering and absorption features of the atmosphere, derived from the properties of its different components (particles, gases), as a function of wavelength and altitude. These optical properties were cast under appropriate forms to be used by radiative transfer (RT) solvers. This model depends on free parameters including surface albedo.

\subsection{Atmospheric properties and composition}

%The vertical structure of the atmosphere is based on HASI data \citep{fulchignoni_etal_2005}, 
%acquired {\it in situ} during the descent of the {\it Huygens} probe.
Most of Titan's atmospheric properties are known from \textit{Cassini} and \textit{Huygens} observations.  The atmosphere temperature and pressure vertical profiles have been measured \textit{in situ} by the Huygens Atmosphere Structure Instrument (HASI) \citep{fulchignoni_etal_2005} during {\it Huygens} probe descent into Titan's atmosphere. The methane mixing ratio has been measured by the Gas Chromatograph Mass Spectrometer (GCMS) onboard \textit{Huygens} \citep{niemann_etal_2005,niemann_etal_2010}. Isotopic ratios of CH$_3$D and $^{13}$CH$_4$ are those published by \citet{Nixon_2012} (see table \ref{isotopes}). For CO, HCN and C$_2$H$_2$ mixing ratios and vertical profiles (see table \ref{isotopes}), we used results derived from CIRS  observations \citep[][also available in VESPA database\footnote{\url{http://vespa.obspm.fr/planetary/data/}}]{vinatier_etal_2015}, extrapolated for the low stratosphere and troposphere, and limited by the saturation vapor pressure.\\

\subsection{Gas spectroscopic absorption}

Molecular spectroscopic data  (table \ref{MDparam}) are mainly provided by HITRAN\footnote{\url{https://hitran.org}} database \citep{rothman_etal_2013}, with a notable exception for CH$_4$ and its isotopes CH$_3$D and $^{13}$CH$_4$ for which we employed the up-to-date theoretical line lists computed by \citet{rey_etal_2017,rey_etal_2018}. These data are publicly available at the Theoretical Reims-Tomsk Spectral database\footnote{\url{https://theorets.tsu.ru} and \url{https://theorets.univ-reims.fr}}  \citep{rey_etal_2016}. Following \citet{DEBERGH2012} for the line broadening, we represented the line profiles 
by a Voigt function with a cut-off at 26 cm$^{-1}$ from the line center. They were then modified with a sublorentzian decay with a wavenumber characteristic length of 120 cm$^{-1}$ for all species except CO. For the latter, we used the specific cut-off recommended by \citet{Lellouch2003}. As VIMS channels wavelengths change through the mission, we followed \citet{clark_etal_2018} recommendations and calculated the spectroscopic absorption consequently.\\

% ----------------------------------------------------------------------------------
\begin{table}[!t]
	\begin{center}
		\caption{\label{isotopes}List of CH$_4$ isotopes ratio taken into account, from \citet{Nixon_2012}, and other absorbing gases mixing ratio derived from CIRS observations \citep{vinatier_etal_2015}.}
		\vspace*{0.5cm}
		
		\begin{tabular}{|l|l||l|l|}
			\hline
			Methane       &     Isotope& Absorbing&Mixing\\
			& ratio& gaz &ratio\\
			\hline
			D/H		& 1,59$\times$10$^{-4}$ &	CO  & 3.3$\times$10$^{-4}$	\\
			$^{12}$C/$^{13}$C		& 86,5&C$_2$H$_2$ & 4$\times$10$^{-4}$	\\
			CH$_3$D/CH$_4$   & 6,36$\times$10$^{-4}$& 		HCN & 3$\times$10$^{-7}$  \\
			$^{13}$CH$_4$/CH$_4$ &  1,156$\times$10$^{-2}$   &&  \\

			\hline
		\end{tabular}
		
	\end{center}
\end{table}
% ----------------------------------------------------------------------------------
\subsection{Aerosol model}
\label{Aerosol model}
\begin{figure}[!t]
	\begin{center}
		\includegraphics[width=6 cm, angle=0]{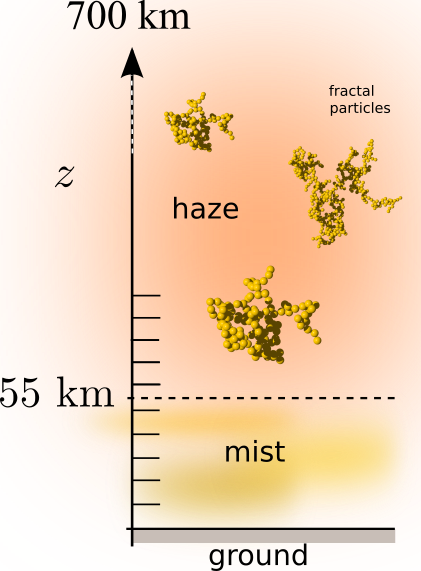}
	\end{center}
	\caption[]{\label{schema_atm}Sketch of the atmospheric profile we assumed in this work. Fractals are from \citet{Botet_1995}, and this illustration if from \citet{Seignovert2019}. According to the adopted global picture, a low altitude layer, extending from the ground to $55$~km, is filled by a misty air containing more or less methane-wet particles. This first layer is topped by a less dense but more extended region of haze settled by fractal particles. In our model, these particles are assumed to be made of $3000$ monomers, each with an individual radius of $50$ nm, the adopted fractal dimension is $\mathcal{D}_f= 2.3$. The single-scattering albedo $\omega_{\rm mist}$ of mist \emph{particles-droplets} has been scaled to the single-scattering albedo $\omega_{\rm haze}$ according to the law: $\omega_{\rm mist}= (0.565 + \omega_{\rm haze})/1.5$ \citep{Doose2016}.}
\end{figure}

Considering the previous works of \citet{tomasko_etal_2008b} and \citet{Doose2016}, we divided the aerosols altitude 
distribution into two layers (see Fig.~\ref{schema_atm}). The upper haze layer extends from the tropopause at 55 km to the top of the atmosphere. It is composed of photochemical aerosols assumed to be fractal aggregates. The lower layer, hereafter called \emph{mist}, is an undefined mixture of condensed material and photochemical haze.

To model the haze optical properties such as the single scattering  albedos and the extinction cross-sections, we adopted the scattering model for fractal aggregate developed by \citet{Rannou1997}. 
The aerosol phase functions that we use are adapted for the backscattering. The preliminary analysis of a sequence of VIMS observations of Titan acquired in November 2012 (flyby T88) led to propose a modification of the aerosol phase function initially recommended by \citet{Doose2016}. During this 5-min long sequence, VIMS acquired 26 consecutive datacubes over the same location, but with varying emission and phase angles. In order to keep the haze opacity constant over the 26 datacubes, one had to change the aerosol phase functions by adding 20\% more of backscattering at all wavelengths. We used these modified aerosol phase functions in the present study. A comprehensive analysis of these observations is out of the scope of the present article, but will be soon subject to a separate dedicated article. 

This model allows us to study different morphology of aggregates, whereas \citet{tomasko_etal_2008b} phase functions are only valid for a restricted set of parameters (\textit{i.e.} a fractal dimension of 2 and 3000 monomers of 50 nm). In our model, we keep the fractal dimension as a free parameter. The aerosols refractive indices were derived from those published by \citet{rannou_etal_2010}. They were adapted to fit the haze single scattering albedo in agreement with the prescription proposed by \citet{Doose2016}. Since mist particles remain poorly known, we made no differences between haze and mist particles phase functions. The haze and mist extinction $k(z,\lambda)$ spectral slopes were obtained with this scattering model, because Doose's results based on the Descent Imager / Spectral Radiometer (DISR) data from \textit{Hyugens} probe are only available between $\sim$ 0.4 and 1.6 $\mu$m.

Consistently, our mist scattering albedo $\omega_{\rm mist}$ follows \citet{Doose2016} prescription, i.e.  $\omega_{\rm mist}= (0.565 + \omega_{\rm haze})/1.5$. Similarly, the transition altitude between "mist scattering regime" and "haze scattering regime"  and the vertical extinction profile $k(z,\lambda)$ came from the same source.

%The single scattering albedo of the mist $\omega_{\rm mist}$ is obtained with \citet{Doose2016} law : $\omega_{\rm mist}= (0.565 + \omega_{\rm haze})/1.5$.   We also follow \citet{Doose2016} for the transition of the single scattering albedo as a function of the altitudes. The vertical extinction profile for the haze and mist layers $k(z,\lambda)$ are given by \citep{Doose2016}. 
The haze and mist optical depths were scaled with respect to the optical thickness obtained by \citet{Doose2016} at 1 $\mu$m. Those profiles were adjusted thanks to a minimization procedure with two free parameters: $F_h$ for the haze, and $F_m$ for the mist, allowing for different ratios for the haze and mist. 
As a consequence, the haze and mist opacity at an altitude $z$ and a wavelength $\lambda$ are:
\begin{equation}
	\Delta\tau_h(z,\lambda)=F_h\Delta\tau_{h_D}(z)\frac{\sigma(\lambda)}{\sigma(\lambda_0)}
\end{equation}
\begin{equation}
	\Delta\tau_m(z,\lambda)=F_m\Delta\tau_{m_D}(z)\frac{\sigma(\lambda)}{\sigma(\lambda_0)}
\end{equation}
with $\sigma(\lambda)$ the extinction cross-section of fractal aggregates, $\Delta\tau_{h_D}(z)$ and $\Delta\tau_{m_D}(z)$ the normalized haze or mist opacity of Doose at 1 $\mu$m for an atmospheric layer at the altitude $z$ times the total opacity of the atmospheric column at $\lambda_0 =$ 1 $\mu$m.

%For a term $n$ of legend polynomial, \textbf{at a given altitude and wavelength}, the polynomial is written :

% \begin{equation}
%\mathcal{P}_n=( \dfrac{\mathcal{P}_{n_a}d\tau_{a}\omega_{a}+ \mathcal{P}_{n_R}d\tau_R}{d\tau_{a}\omega_{a}+d\tau_R})\dfrac{1}{\mathcal{P}_0} 
%\end{equation}
%With $ \mathcal{P}_n$ a legend term $n$ of the phase function used in the RT model,  $\mathcal{P}_{n_a}$ a legend term  $n$ of the phase function of the aerosols, $\mathcal{P}_{n_R}$ a legend term $n$ of the phase function due to Rayleigh scattering, $d\tau_{a}$ the aerosol opacity, $\omega_{a}$ the aerosols' single scattering albedo, and $d\tau_R$ the opacity due to Rayleigh scattering in atmospheric methane. 

%Voight (convolution d'un profile laurentzien et d'une gaussienne)
%J'ai mis le cut-off a 26 cm-1  du centre de raie avec une décroissance  sublorentzien de 120 cm-1. C'est la prescription de de Bergh et al. (2012).

\subsection{Other components and technical aspects}

The scattering coefficients due to Rayleigh scattering were derived from equations  (2.30) and (2.31) of \citet{Hansen1974}. They adapted to Titan's atmosphere using the CH$_4$ column density. Our model also includes N$_2$-N$_2$ and  N$_2$-H$_2$ collision-induced absorption (CIA) \citep{Lafferty1996, mckellar1989}. Titan's surface is assumed to scatter as a lambertian surface.
\\

We discretized the atmosphere into $70$ layers uniformly distributed from ground level up to an altitude of $700$~km, since it was found as an acceptable compromise between accuracy and computation time \citep{mckay_etal_1989,Cours2020}. In addition, for each layer we computed the average optical depths, phase functions and single scattering albedos for all VIMS-IR channels.
We have the choice between two solvers to solve numerically the radiative transfer equations: \texttt{SHDOMPP} developed by K. Evans\footnote{\url{http://coloradolinux.com/shdom/shdomppda/}} \citep{evans_2007} that runs in a plan-parallel approximation, and \texttt{SPSDISORT} from the library libRadtran\footnote{\url{http://www.libradtran.org/doku.php}} \citep{mayer2005, libRadtran2016} that includes a pseudo-spherical geometry. The first solver was found to compute ten times faster than%reduce the computation time by a factor of ten compared to the execution time of
the second, which has a wider validity range since it includes spherical corrections. The optical depths of the different gases were calculated according to the correlated-k approximation \citep{goody_etal_1989}, including four terms for each channel to represent the opacity of the gases.

In radiative transfer models such as \texttt{SHDOMPP}  or \texttt{SPSDISORT}, the phase functions have to be expanded on a basis of  Legendre polynomials, and we use 100 terms. In the model, radiance is calculated for a given number of directions (that is, streams). The resolution and computation time increase with the number of streams. The section \ref{stream_section} treats in details the optimal number of streams to use in the model.

The synthetic spectra were fitted to observational data in the bands using the Levenberg-Marquardt algorithm \citep{levenberg_1944,marquardt_1963}.
For that purpose we employed the implementation of this method available under the form of the \texttt{LMDIF} subroutine, 
part of the \texttt{MINPACK} library \citep{more_minpack} written in \texttt{FORTRAN}.\\

%The final phase function depends on the haze and mist phase function, the opacity and simple scattering albedo of the aerosols, Rayleigh scattering phase function and opacity. 

%
% ----------------------------------------------------------------------------------
\begin{table}[!t]
	\begin{center}
		\caption{\label{MDparam}List of gaseous species accounted for in our radiative transfer model.}
		\vspace*{0.5cm}
		
		\begin{tabular}{|l|l|l|}
			\hline
			Species       & Instrument    & Spectrum\\
			
			\hline
			CH$_4$        &  GCMS & TheoReTS  \\
			CH$_3$D       & CIRS/GCMS & TheoReTS \\
			$^{13}$CH$_4$ &  CIRS/GCMS & TheoReTS \\
			CO            &  CIRS &HITRAN \\
			C$_2$H$_2$    &  CIRS & HITRAN\\
			HCN           &  CIRS &HITRAN \\
			\hline
		\end{tabular}
		
	\end{center}
\end{table}
% ----------------------------------------------------------------------------------
%
%
%===================================================================================================================================
%===================================================================================================================================
%

%
%===================================================================================================================================
%\begin{figure}[!t]
%\begin{center}
%\includegraphics[width=8.5 cm, angle=0]{fig/example_fit_spectrum_1.eps}
%\end{center}
%\caption[]{\label{example_fit}(a) Infrared spectra (black circles) extracted from the pixel $(3, 19)$ of VIMS cube \texttt{V1809727868}, compared 
%           to the synthetic spectra (red solid lines) computed with our radiative transfer model. (b) The corresponding surface albedos, derived for 
%          each atmospheric window. Vertical dashed lines only mark the boundaries of windows.}
%\end{figure}
%===================================================================================================================================
%

%
%===================================================================================================================================

%===================================================================================================================================
%
% 
%%%%%%%%%%%%%%%%%%%%%%%%%%%%%%%%%%%%%%%%%%%%%%%%%%%%%%%%%%%%%%%%%%%%%%%%%%%%%%%%%%%%%%%%%%%%%%%%%%%%%%%%%%%%%%%%%%%%%%%%%%%%%%%%%%%%
%%  :

\section{On the error of VIMS}
\label{error}

The photometric uncertainties in VIMS calibrated cubes are not well known, and depend on the considered channel \citep{brown_etal_2004}. \citet{Sromovsky2011} established a global average error of 2.5\% of the $I/F$ values, with an $I/F$ offset error of 5$\times$10$^{-4}$. In order to provide a more accurate description of the instrument noise, we developed a new method based on the analysis of inter-pixels variability. With this method we can estimate combined errors due to the instrument, data calibration and post treatment. Assuming a slow pixel to pixel variation, we perform an estimate of the intrinsic variability of adjacent pixels. 

First, for a given data cube, we chose about 10 pixels evenly distributed over the image. We took into account the 8 first neighbors around them, forming a 3$\times$3 box around the selected pixel. For each set of 9 pixels, we calculated at each wavelength %band -- as there is no influence of the surface --  
the mean $\overline{I/F}$ and standard deviation $\sigma$. %defined respectively by :
%\begin{equation}
%\overline{I/F}=\frac{1}{N} \sum_{n=1}^{N}(I/F_n)
%\end{equation}
%\begin{equation}
%sd=\sqrt{\dfrac{1}{N-1}\sum_{n=1}^{N}(I/F_n-\overline{I/F})^2} 
%\end{equation}
%Where N is the total number of pixels, here N=9. 
Finally we derived an estimator of the relative standard deviation $\Sigma$ of each 9-pixels set: 

\begin{equation}
	%\Sigma={\sigma}/{\overline{I/F}}
	\Sigma = \text{std}(I/F) / \text{mean}(I/F)
\end{equation}

\begin{figure}[h]
	\begin{center}
		\includegraphics[width=16 cm, angle=0]{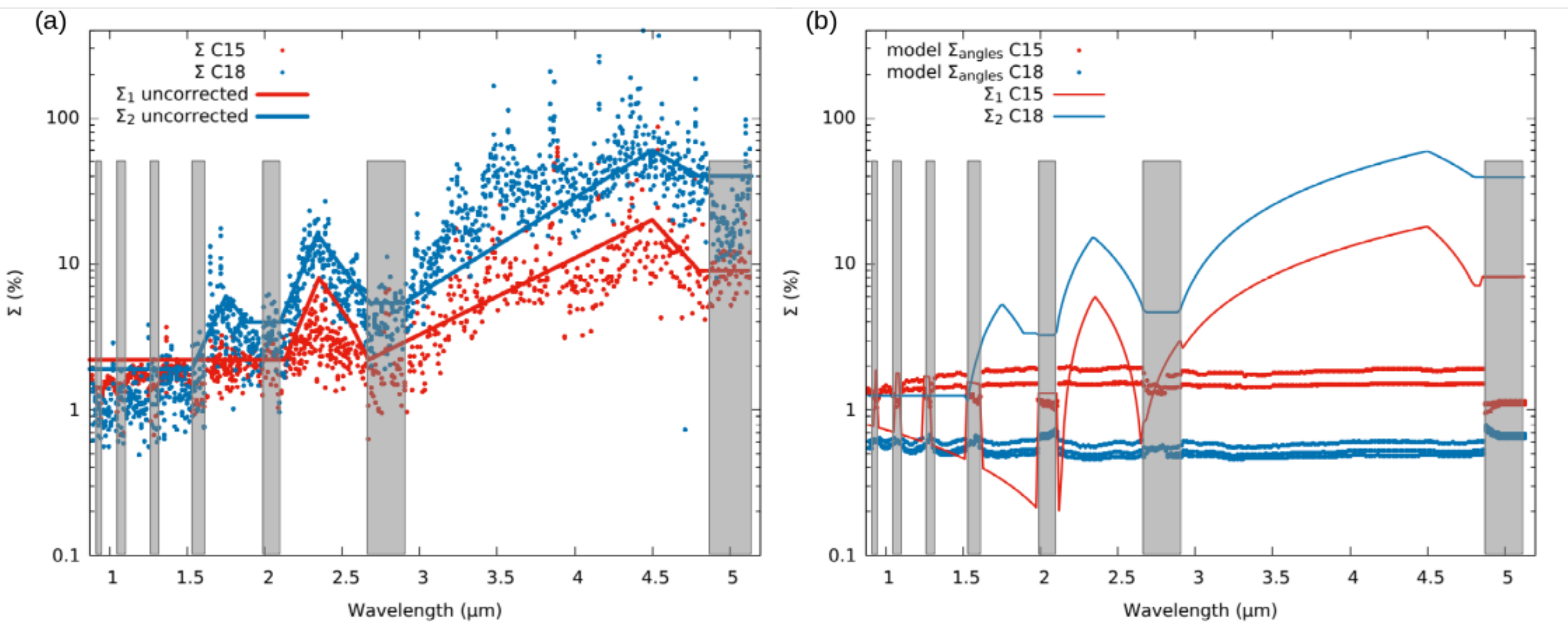}
	\end{center}
	\caption[]{\label{sd_1}\label{sd}(a) Chosen spectral profiles of $\Sigma$ (lines), and retrieved values of $\Sigma$  (points) for C15 (red) and C18 (blue). Titan's atmospheric windows are in gray rectangles. (b) Final spectral profiles of $\Sigma$ (lines) for the C15 ($\Sigma_1$) and C18 ($\Sigma_2$), where we subtracted from our previous uncorrected profiles the $\Sigma$ due to the angle variation (dots).}
\end{figure}

For C15 and C18, we calculated $\Sigma$ for the whole spectrum as we need it to estimate the surface albedo error bars. The results are represented by dots in Fig.~\ref{sd_1}-a. The method is only strictly valid in the bands because in the windows we get a variability of the signal related to the natural variability of the surface. We calculated $\Sigma$ in the atmospheric windows for comparison purpose. The geometry changes between each pixels and can influence the retrieved $\Sigma$. This will be corrected by another step of the method.

By analyzing different set of pixels, over a few images, we can first notice in Fig.~\ref{sd_1}-a that $\Sigma$ depends on the cube and the wavelength. 
We determined specific profiles of $\Sigma$ for each image of interest. They were chosen manually, following the general shape of the results ($\Sigma_1$ and $\Sigma_2$ uncorrected in Fig.~\ref{sd_1}-a for C15 et C18 respectively). We chose to cap $\Sigma$ below 2.5 $\mu$m and at 5 $\mu$m by a high value so we do not underestimate the error bars. To obtain the uncorrected $\Sigma$ in atmospheric windows, we interpolate the uncorrected $\Sigma$  of adjacent bands.

We then determined the error due to variations in viewing geometry from one pixel to an other. We obtained it with the same method, but instead of the cube spectra, we used synthetic spectra with fixed parameters for the surface albedo and the aerosols. For these synthetic spectra, only incidence and emergence angles change, similarly as those of the cube. In that way, the only differences in $I/F$ between pixels of the same cube are due to angle variations. With the differences in incidence and emergence angles between C15 and C18, the thickness of the atmosphere layer crossed by the beams is greater for C18 than for C15. As a result, contrary to C18 where incident angles are higher, the error due to angle variations is lower in the bands for C15.

We subtracted from our previous profiles the $\Sigma$ due to angle variation. Final profiles of $\Sigma$ are shown in Fig \ref{sd}-b. Here $\Sigma_1$ is adapted for C15, whereas the profile $\Sigma_2$ is adapted for C18. 
Theoretically, in the bands we should have a negligible influence from the surface. 
In Fig.~\ref{sd_1}-a for C15 and C18 (red and blue points) we tend to have a lower $\Sigma$ in the windows. VIMS uncertainties depend on the number of photons. The greater $I/F$, the smaller uncertainties. Because of the interpolation of $\Sigma_{1,2}$ uncorrected for the windows, we may overestimate the error in the atmospheric windows where we have a greater $I/F$ than in the bands.

The corrected $\Sigma$ for C15 and C18 in Fig.~\ref{sd}-b show the influence of incident and emergent angles. The corrected $\Sigma$ for C15 is higher in the windows than in the bands, under 2 $\mu$m. For C18 it is slightly lower, except at 5 $\mu$m.
Generally, there is an enhanced variability and error around 2.35 $\mu$m, and beyond 3 $\mu$m. Our results there are quite different from the 2.5\% estimated by \citet{Sromovsky2011}, but below 2$\mu$m the values are of the same order in the windows. The influence of the temperature inside VIMS, causing IR radiation, is greater at large wavelengths. That could partly explain the increasing uncertainties with the wavelength. The low number of photons at greater wavelength is also another factor.

These $\Sigma$ are significant in the bands up to 2.5 $\mu$m for the calculation of the $\chi^2$ and the minimization of the function $\mathcal{Y}$ used to retrieve atmospheric parameters F$_h$ and F$_m$, the scaling factors for the haze and the mist:
\begin{equation}
	\mathcal{Y}_n= \dfrac{|(I/F)_{n}^{obs}-(I/F)_{n}^{mod} |}{\Sigma_n \times(I/F)_{n}^{obs}}
\end{equation}
\begin{equation} \label{chi2}
	\chi^2= \frac{1}{N}\sum_{n=1}^{N}\mathcal{Y}_n^2
\end{equation}
with $N$ the number of channels we used to determine the parameters, $I/F_{obs}$ the observed $I/F$ and $I/F_{mod}$ the modeled $I/F$. $\Sigma_{\lambda}$ is also used later to estimate the albedo error bars.

Once surface albedos are determined for each channel, we can finally evaluate the uncertainties of the surface albedo for a wavelength $\lambda$. We need the values of $\Sigma_{\lambda}$ previously calculated and the derivative of the albedo $A_\lambda$ with respect to  $I/F_\lambda$:

\begin{equation}\label{err}
	\sigma_{A_{\lambda}}=\left|\dfrac{dA_{\lambda}}{dI/F_\lambda}\right|\Sigma_{\lambda}
\end{equation}

%We can then get $\sigma_A$ the absolute error of the albedo :

%\begin{equation}
%\sigma_A=\dfrac{\sigma_{A^+}-\sigma_{A^-}}{2}
%\end{equation}

These errors will be used to calculate our results uncertainties.

\section{Test and preparation of the model for surface analysis}
%{\color{Violet}Faire une partie complète, montrant les tests à Df = 2.4 et 2.3 par rapport à 2, et montrer le meilleur résultat, avec des figures de $I/F$ et d'albédo de simple diffusion
%Parler des tests techniques et de la préparation du modèle -> ajustement de la brume, le nombre de stream utilisés, la dimension fractale}
%----------
In this section we present sensitivity tests performed with the radiative transfer model, using \texttt{SPSDISORT} and \texttt{SHDOMPP}. The scope of this section is to define the number of stream used to discretize the intensity field in the model and to define the limit incident and emergent angles for using the parallel plane model beyond which we must account for the atmosphere sphericity. Then, we also describe how we set the atmosphere free parameters and our observed data fitting method.

\subsection{Tests of the model against the number of streams}
\label{stream_section}
%{\color{Violet} Nombre de Strings et Streams, figure montrant à partir de combien de strings et streams %on peut estimer que le modèle ne donne plus de résultat différent (shdompp et spsdisort)

%	* Definition of a testing case (Fh=1,Fc=1, mu1, mu0, phi + As=?)

%	* Criterion for stability of the result (e.g, we want to reach
%	stable results within 1\%, 0.1\%, 0.01\% for each wavelength as
%	NMU*NPHI increases ? Or 1\%, 0.1\% in average ? Or at some
%	wavelength ???)

%	* Test on the number of streams with the SHDOMPP + SPSDISORT
%	(Q1: how much NMU and NPHI to use for reaching a stable $I/F$ at
%	ALL wavelengths Q2: Test on NKS (correlate-k) )

%	* Conclusion (which value of NMU and NPHI for each model, which
%	value of NKS ?)}

%Both solvers have different computational time. SHDOMPP is faster, but is limited with high incident angles. SPSDISORT is more accurate for high incident angles, but its computational time is slow.
A solver for the radiative transfer equations calculates the intensity field in a finite number of directions for each atmospheric layer. This is what we call the number of streams.
In order to optimize the model execution time, we tested the solvers accuracy with different number of streams. A high number of streams increases the accuracy, but also the execution time.

We define a testing case with fixed haze and mist parameters $F_h = F_m=1$. The surface albedo is set at a uniform value of 0.05.  We use a near-nadir viewing with an incidence angle of 3$^{\circ}$, an emergence angle of 5$^{\circ}$ and a phase angle of 7$^{\circ}$. Those are extreme conditions for the \texttt{SHDOMPP} because the nadir configuration requires more accuracy, i.e. a higher number of streams than other geometries \citep{evans_2007}. 

\begin{figure}[h]
	\begin{center}
		\includegraphics[width=8 cm, angle=0]{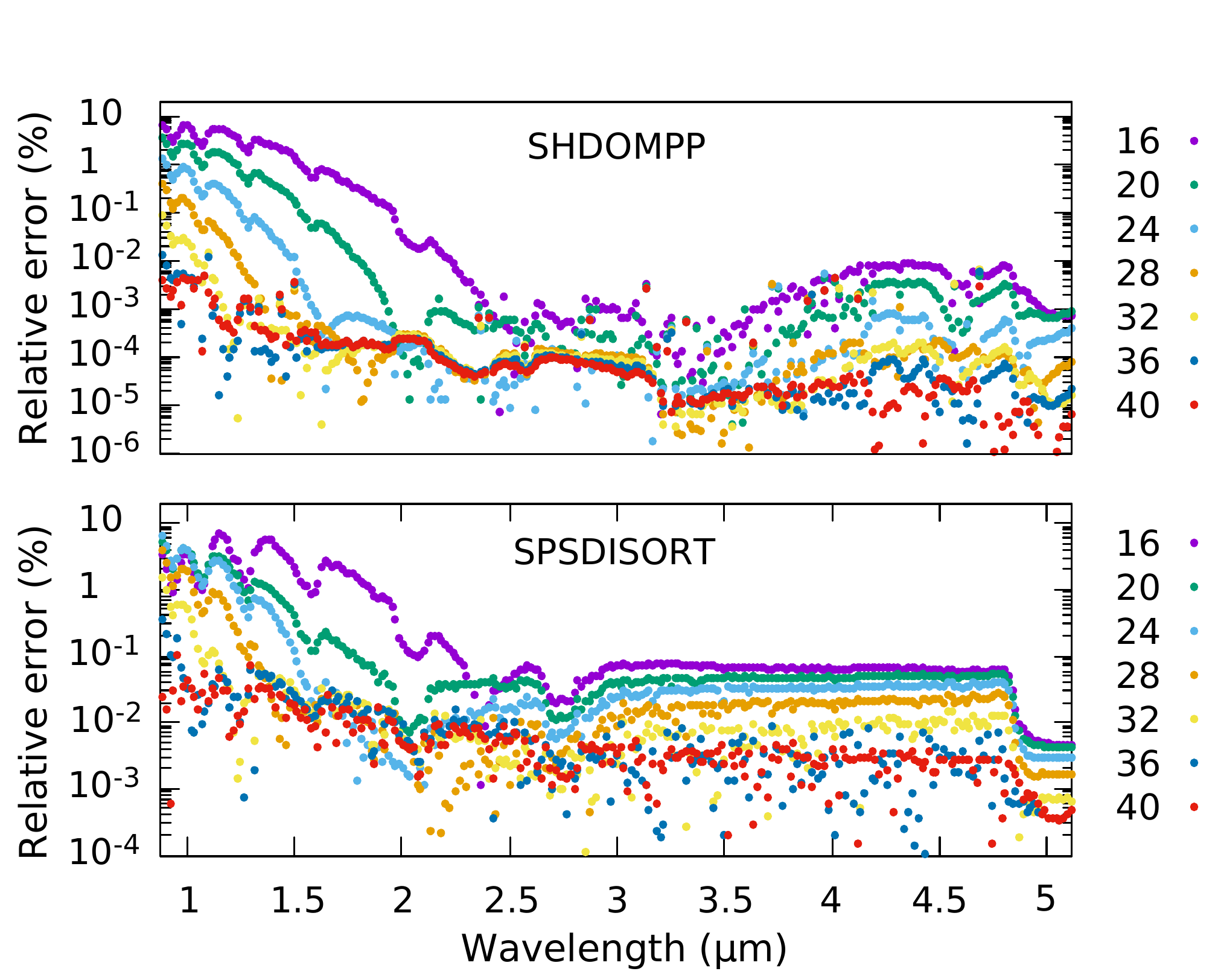}
	\end{center}
	\caption[]{\label{streams}Differences in \% between the $I/F$ calculated with \texttt{SHDOMPP} (top panel) and \texttt{SPSDISORT} (bottom panel) with different number of streams compared to the results with 100 streams. Incidence: 3$^{\circ}$, emergence: 5$^{\circ}$, phase: 7$^{\circ}$. The relative error is: $\Delta r = |\frac{I/F_{Nstr}-I/F_{100}}{I/F_{100}}|$ .}
\end{figure}

We set a reference with a simulation made with 100 streams, as we would never use a number so high due to the runtime, even if the accuracy of the results is better. We then compare the model performance with different number of streams. In order to estimate the minimum number of streams required to have less than 0.5\% difference in $I/F$ at each wavelength with the reference case, we start with 16 and we increase numbers by steps of 4 streams (Fig.~\ref{streams}). As the differences between the reference case and the tests decrease, the differences between the tests also decrease, allowing us to choose an optimum number of stream.

For the solver \texttt{SHDOMPP},  we have less than 1\% difference when using more than 28 streams. With 32 streams the difference falls below the 0.5\% limit. The solver \texttt{SPSDISORT} requires a minimum of 36 streams to have less than 0.5\% difference compared to the 100-streams test case. In view of the results in Fig.~\ref{streams}, and taking into account the geometry and the increase of computation time with the number of streams, we decided to use for \texttt{SHDOMPP} a number of 32 streams up to 1.7 $\mu$m, and 24 streams for the rest of VIMS-IR wavelengths. For the \texttt{SPSDISORT}, we work with 40 streams up to 1.7 $\mu$m, and 24 streams for the reminder of the spectrum. Doing so, we have less than 0.5 \% of difference with the reference test at each wavelength, with a smaller computation time. For example, for a cube made of 4096 pixels, the RT model using \texttt{SPSDISORT} has a runtime of less than 24 hours when using 500 processors. 

\subsection{Tests of the model in a spherical viewing}
%{\color{Violet}Mettre une justification sur l'utilisation de SHDOMPP vs. SPSDISORT en s'appuyant sur la Fig.~\ref{ene_incid} et eventuellement
%	un exemple de comparaison d'albedos calcules avec SHDOMPP/SPSDISORT.}\\

\begin{figure}[h]
	\begin{center}
		\includegraphics[width=8.5 cm, angle=0]{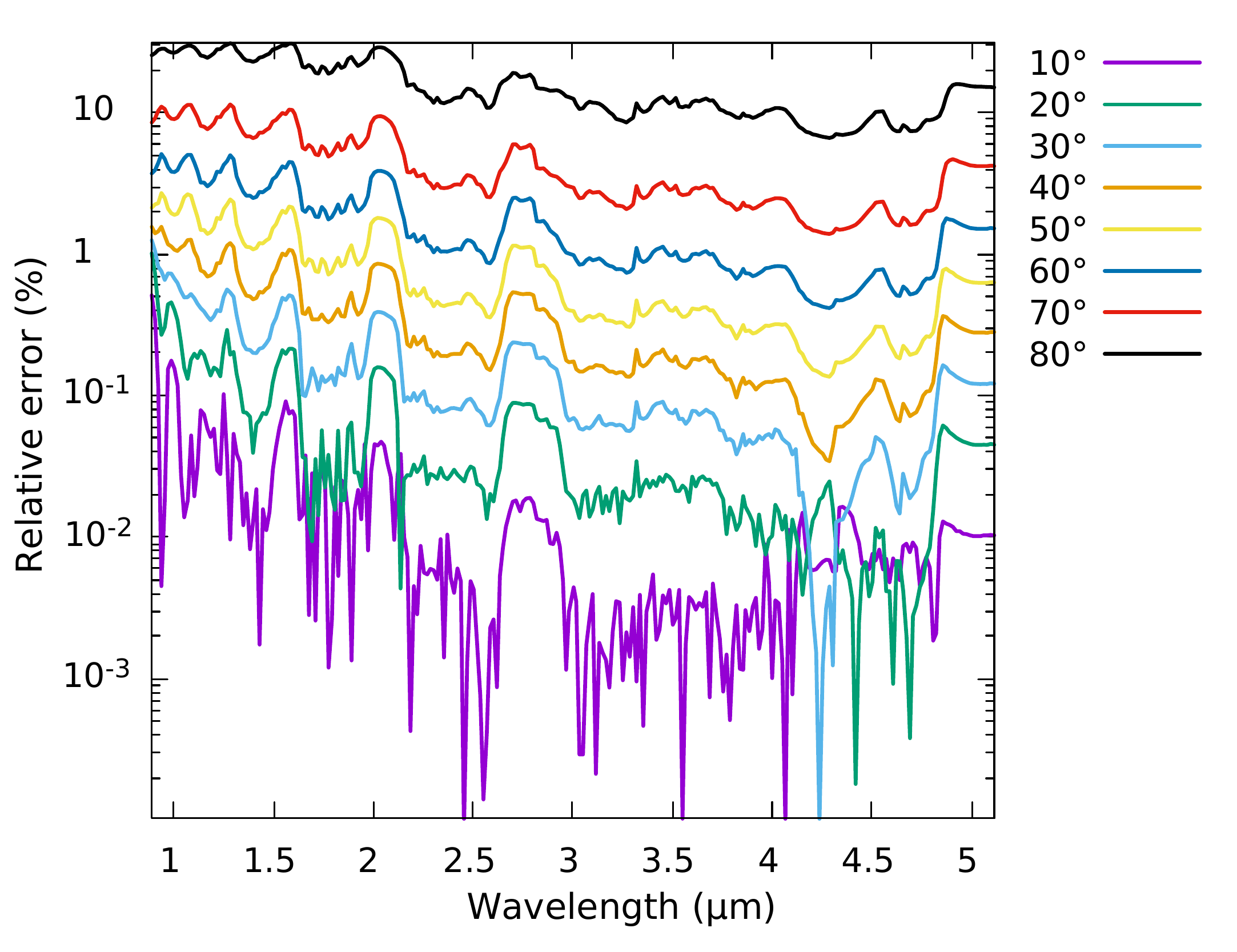}
	\end{center}
	\caption[]{\label{shdompp_spsdisort}Differences (\%) between $I/F$ calculated with \texttt{SHDOMPP} and \texttt{SPSDISORT} for different incidence angles. Emergent angle: 10$^{\circ}$. Relative error: $\Delta r =  |\frac{I/F_{SHD}-I/F_{SPS}}{I/F_{SPS}}|$}
\end{figure}

\citet{Barnes2018}  consider as acceptable the use of parallel-plane approximation up to 60$^{\circ}$ for incidence or emergence angles. In this work, we compared the results obtained respectively with \texttt{SHDOMPP} in a parallel-plane approximation and \texttt{SPSDISORT} in a pseudo-spherical geometry at different incident angles (see Fig.~\ref{shdompp_spsdisort}). For this testing case, we use 32 streams for both models, and a emergence angle of 10$^{\circ}$. We change the incidence angle from 10$^{\circ}$ to 80$^{\circ}$ with a 10$^{\circ}$ increment.

Fig.~\ref{shdompp_spsdisort} shows that for incidence angles lower than 30$^{\circ}$, we have differences between solvers smaller than 1\%, which is in agreement with \citet{Barnes2018} result with a Monte-Carlo RT model. Errors reach more than 5\% in Titan's atmospheric windows at 60$^{\circ}$. 
%To evaluate the differences with respect to VIMS uncertainties, we calculated the standard deviations of several squares of 9 pixels on the images. We found that VIMS error increases form 3,5\% at short wavelengths to 15\% at least beyond 3 $\mu$m. % .We do find an increase of error in the bands at larger wavelength.
%We need to be conscious that the SPSDISORT is a pseudo-spherical model, and that one of the limits is that the emergence angle is not well taken into account at high values.
Giving these results, we decided for our study to employ \texttt{SHDOMPP} with C15 (Fig.~\ref{ene_incid}-a) when the incident angles are below 50$^{\circ}$. For C18 (Fig.~\ref{ene_incid}-b), we chose to work with \texttt{SPSDISORT}, as in our region of interest incidence angles go beyond 55$^{\circ}$.

%Differences appear starting from 30$^{\circ}$ angles which is in correlation with \citet{Barnes2018}.
% They are most visible in the first window, but it stays within VIMS error bars calculated from the standard deviation of multiple 9-pixels squares spectra, and this windows is mostly unused. We decided to keep using the plan-parallel approximation.

%----------
%* Definition of a testing case (Fh=1, Fc=1, mu1, mu0, phi + As=?)
%How we make mu0, mu1 and phi varying ? Do we test all cases ?
%* Criterion for stability of the result.
%* Test on the various angles ?
%* Conclusion (in which angle interval we can use safely the
%SHDOMPP ?)

\subsection{Haze and mist properties for surface analysis}
The last step of the radiative transfer model setting consisted in defining the haze and mist properties that best reproduce the spectra in the bands, where surface has no influence. With our model, we retrieved haze and mist properties by fitting a spectrum taken in a dark tropical area of Titan in order to minimize the influence of surface. More specifically, we chose 16 test pixels where surface albedo is known to be extremely dark \citep{griffith_etal_2012}, from the cube v1567239055, acquired on August, 31, 2007. In that way, we have set the model in such a way a negative value of the surface albedo is not possible even where surface albedo is especially low. VIMS uncertainties defined in section \ref{error} were taken into account in the algorithm.

\begin{figure}[h]
	\begin{center}
		\includegraphics[width=8.5 cm, angle=0]{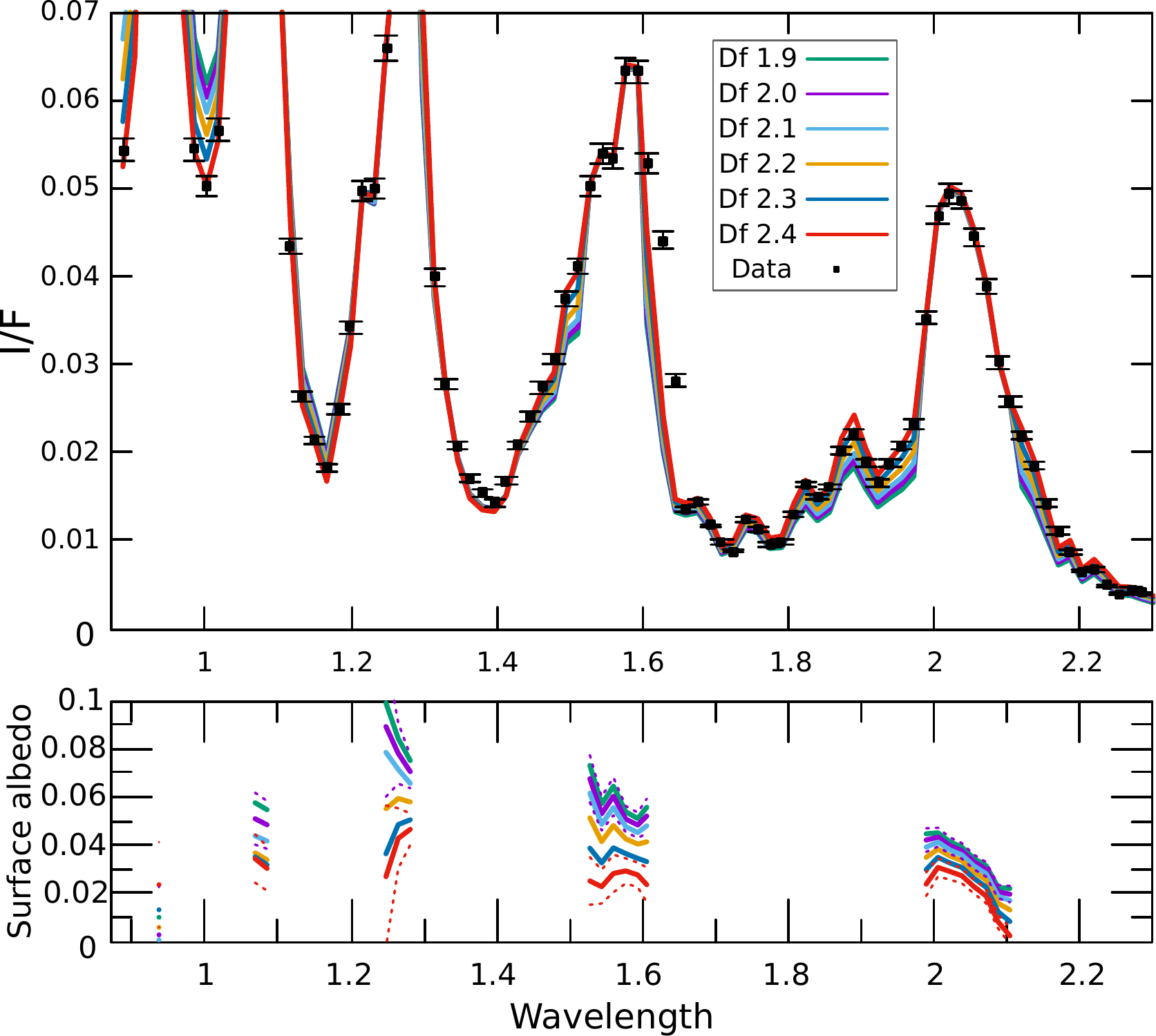}
	\end{center}
	\caption[]{\label{Df}Fits of the pixel (38,19) of the cube v1567239055, using different fractal dimension (top), and the corresponding surface albedo (bottom). Dotted lines: uncertainties for  $\mathcal{D}_f$ = 2  and  $\mathcal{D}_f$ = 2.4.}
\end{figure}

We used the standard case for aerosols morphology described in section \ref{Aerosol model}.  In Fig.~\ref{Df} with a fractal dimension of $\mathcal{D}_f$=2 we obtain a good fit except for the band around 1 $\mu$m. This is a recurring problem for each VIMS cube.
To solve this problem we first performed several tests, using other tholins refractive index,  varying the single scattering albedo with a constant factor, or using the condition F$_h$=F$_m$. However this did not improve the fit around 1 $\mu$m. We finally tested other values of fractal dimension.

\begin{figure}[h]
	\begin{center}
		\includegraphics[width=8.5 cm, angle=0]{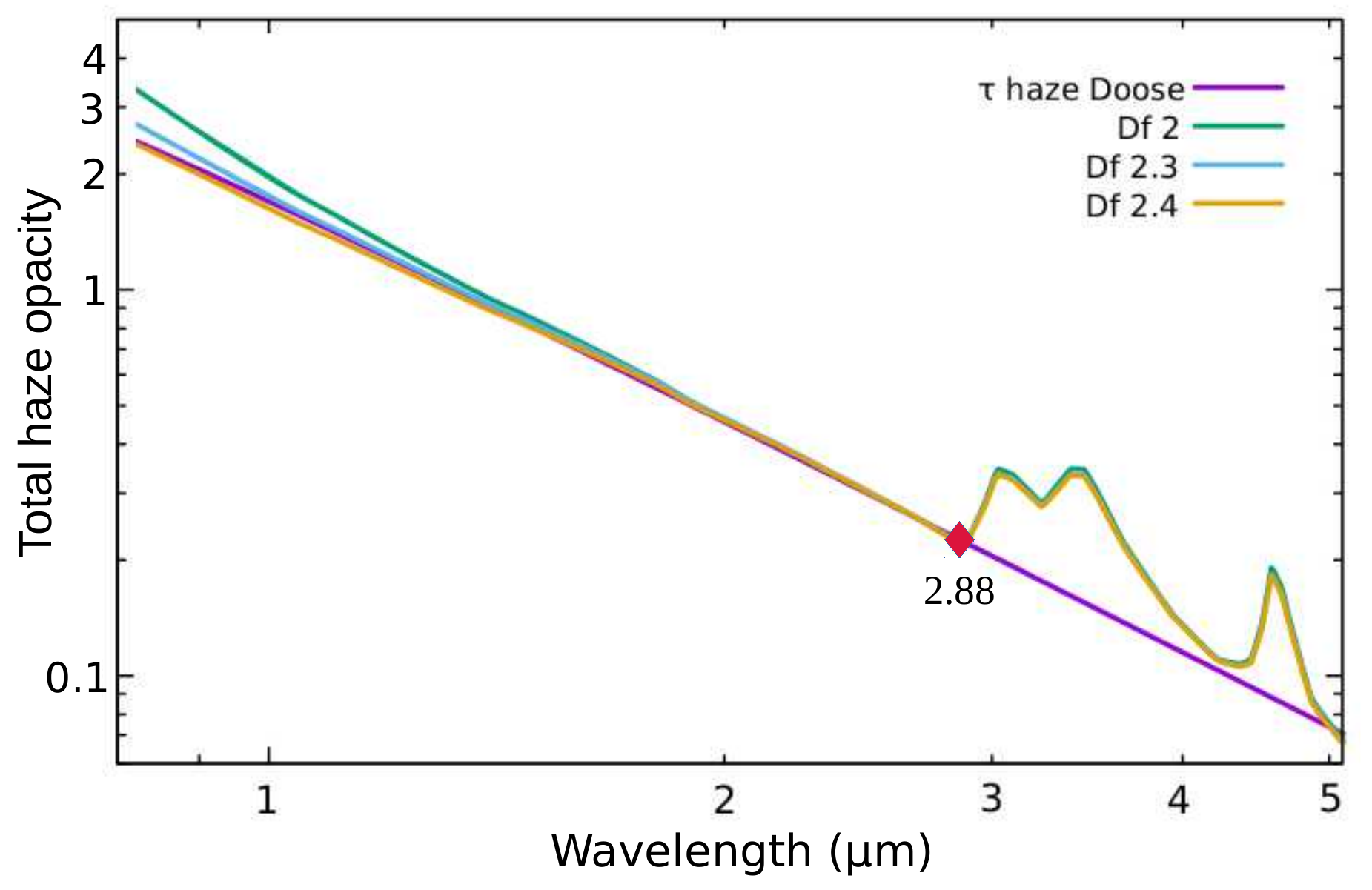}
	\end{center}
	\caption[]{\label{tau}Total haze opacity with F$_h$ and F$_m$ equal to 1, and angles of incidence and emergence at 10$^{\circ}$, with different values of the fractal dimension. The curves follows the microphysics aerosol model from \citet{rannou_etal_2004}. Results are adjusted to fit \citet{Doose2016} curve at 2.88 $\mu$m}
\end{figure}

Varying  $\mathcal{D}_f$ has an impact on the spectral slopes of the haze and mist optical depths (see Fig.~\ref{tau}), and on the phase functions. We kept the number of monomers at 3000 since it was defined from the solar aurea with DISR, which depends in first place on the amount of grains, regardless the fractal dimension.

Then, in order to retrieve the surface albedo and the $\chi^2$ (see Eq. \ref{chi2}) corresponding to each $\mathcal{D}_f$, the parameters F$_h$ and F$_m$ are set free to fit the data in the bands. We found that the fit is greatly improved with $\mathcal{D}_f >$ 2.0 in each of the selected pixels at short wavelengths (see Fig.~\ref{Df} for one of the test cases), especially for the band around 1 $\mu$m, and between 1.8 and 2 $\mu$m. The $ \chi^2$ is generally reduced by a factor of 2 between $\mathcal{D}_f$=2 ($\chi^2=250$ in the example) and $\mathcal{D}_f$=2.3 ($\chi^2=120$ in the example). We have obtained a better fit at 1 $\mu$m, at the short wavelength of the 1.6 $\mu$m window, and around 1.8-2.0 $\mu$m with $\mathcal{D}_f=2.3-2.4$, than with $\mathcal{D}_f=2$ as proposed by \citet{tomasko_etal_2008b}. The surface albedo is also impacted with the change of the fractal dimension. The slope of the surface albedo at 1.28 $\mu$m for $\mathcal{D}_f > 2.1$ is reversed and the shape at 1.55 $\mu$m is flatter even if the graphic scale tends to smooth the variations. Lastly, the surface albedo globally decreases when increasing fractal dimensions, except at 0.93$\mu$m (Fig.~\ref{Df}-bottom).

To compare the total haze opacity with \citet{Doose2016} analytical model, we have chosen a test case with $F_h = F_m = 1$. The optical depths obtained in Fig.~\ref{tau} for different fractal dimensions are adjusted for a better viewing. The differences of $I/F$ at 1 $\mu$m in Fig.~\ref{Df} are correlated with the variations of haze opacity $\tau$ as a function of wavelength in Fig.~\ref{tau}: the spectral slopes are a close equivalent to the one recommended by \citet{Doose2016} beyond $\sim$ 1.3 $\mu$m. Below this threshold, the smaller is $\mathcal{D}_f$, the steeper is the spectral slope. While opacity remains the same above 1.3 $\mu$m, fractal aggregates lead to a too large opacity at 1 $\mu$m for $\mathcal{D}_f$ = 2.0, and yield an opacity comparable to the expected opacity at $\mathcal{D}_f$ $\sim$ 2.3 - 2.4. As Fig.\ref{tau} shows, $\mathcal{D}_f$ = 2.4 is a good candidate, but below 0.8 $\mu$m the result is less consistent with \citet{Doose2016} than for $\mathcal{D}_f$ = 2.3.

These results imply that aerosols are more compact than expected at least, at this location and date. They also imply that the fractal dimension value is between 2.3 and 2.4. This is the case for C15 and C18 as well, as the fit is also improved for the pixels we have chosen to do the comparison with. \\

Based on these results, we decided for our study to employ a fractal dimension of 2.3, even if we obtain almost similar results with $\mathcal{D}_f$ = 2.4. As seen in Fig.~\ref{Df}, with $\mathcal{D}_f$ = 2.4  the fit is the best at 1 $\mu$m, but $\chi^2=150$ which is larger than for $\mathcal{D}_f$ = 2.3 ($\chi^2=120$), meaning that the fit is not as great at other wavelengths. %Both $\chi^2$ may seem high, but it is not unusual to have a $\chi^2$ greater than 1 using Eq. \ref{chi2}, because both denominators $\Sigma$ and $(I/F)^{obs}_n$ are often lower than 0.05 in the bands.

%----------
%* Test with the standard choice for haze and mist (cf SECTION 3).
%Choix d'un pixel test pour calibrer le modèle.
%Here we show the impact of the fractal dimension on the opacity, using tholins refractive index derived from \citet{rannou_etal_2010}. We chose a test case with F$_h$ and F$_m$ = 1 in order to keep Doose's analytical model unchanged for the comparison. The results obtain with different fractal dimensions are scaled for a better viewing.

%The slopes  with different $\mathcal{D}_f$ fit well Doose's results between 1.2 and 2.5  $\mu$m. The differences appears below 1.2 $\mu$m, where the slope for $\mathcal{D}_f$=2 is too steep. With $\mathcal{D}_f$=2, the opacity increases too fast at short wavelength compared to Doose's analytical results. Consequently the bands at short wavelength can not be well fitted. Below 1.2  $\mu$m only cases with  $\mathcal{D}_f\geq$2.3 seems compatible with \citet{Doose2016}. We did not change the number of monomers, but adapted the aerosol optical constants to conserve the same single scattering albedo as Doose. 

%- test sur
%une surface sombre (il me semble que c'est pour ça qu'on a utilise
%le lac tropical - pour voir comment éviter des As < 0)?

%Choice for the mist (how it is related to the haze ? different
%options ?)

%Conclusion : plot tau$_h$, tau$_m$, ssa$_h$, ssa$_m$ with wavelength ?
%Montrer des fits types...
%{\color{Violet} Dimension fractale de la brume}

\subsection{Tests of the final model on \textit{Huygens} landing site}
Thanks to \textit{Huygens} probe, the aerosol vertical profile is known at \textit{Huygens} landing site i.e., at 192.3°W 10.5°S
\citet{Kazeminejad2011} on January, 14, 2005.  To confirm the model validity, we ran a test on a cube containing \textit{Huygens} landing site, and at a date close to \textit{Huygens} landing day. For cubes respecting these criteria, the retrieved values of F$_h$ and F$_m$ should be close to 1 as we use the vertical extinction profile retrieved from \textit{Huygens} given by \citet{Doose2016}. %Even more so if we impose $F_h = F_m$. 
The tests where made with our final model settings, $\mathcal{D}_f=2.3$ and the stream number chosen in section \ref{stream_section}. Incidence and emergence angles are in the validity domain of \texttt{SHDOMPP}.

\begin{figure}[h]
	\begin{center}
		\includegraphics[width=8.5 cm, angle=0]{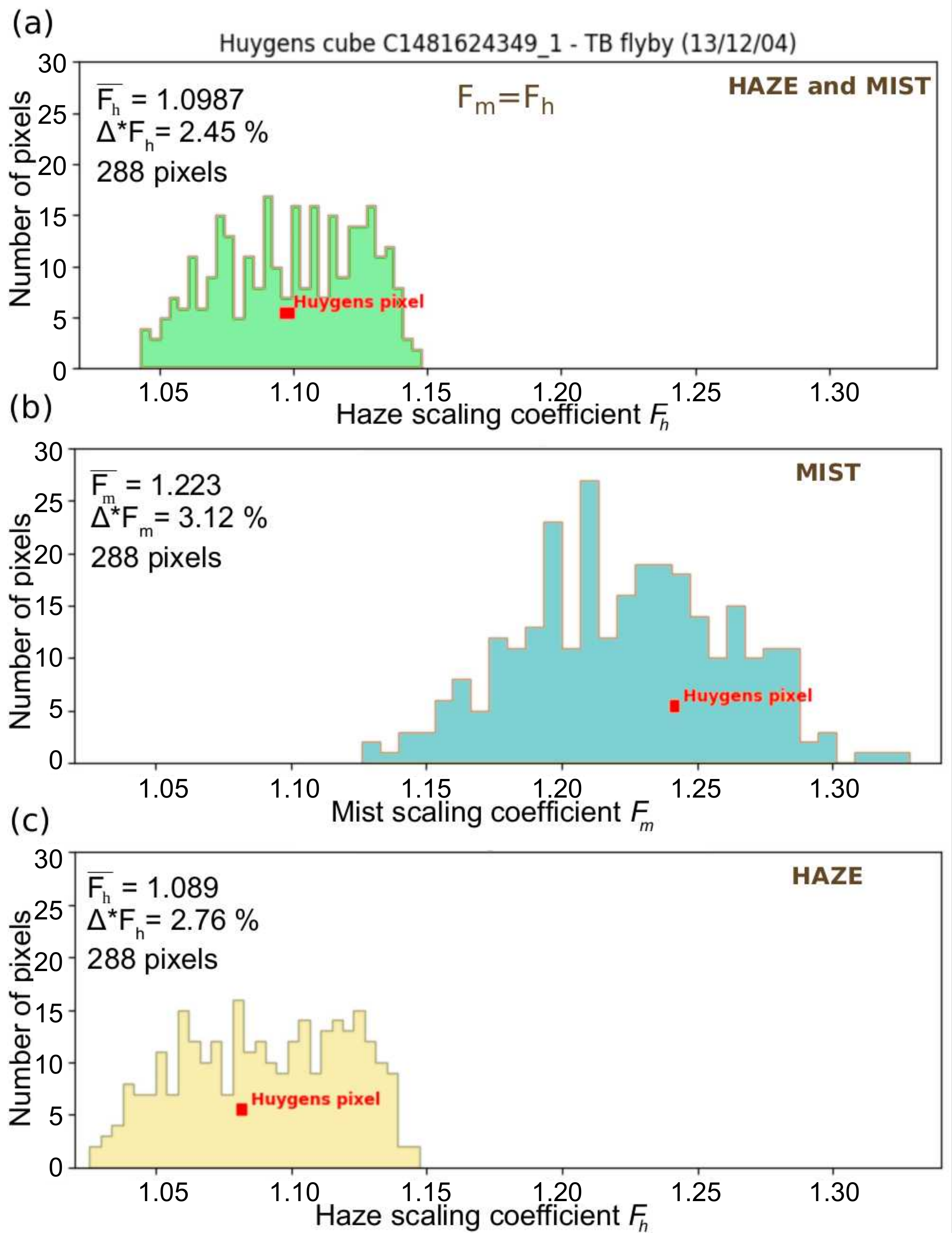}
		
	\end{center}
	\caption[]{	\label{C14FC=FH}\label{C14FCdiffFH}(a) Histogram of retrieved haze and mist parameters for $F_h = F_m$ for the cube C14. (b,c) Histogram of retrieved haze and mist parameters respectively for the cube C14 for the case $F_h \neq F_m$. Results at \textit{Huygens} landing site are indicated in red. The relative standard deviation is noted $\Delta F_h$* in order not to be confused with $\Sigma$. (b)}
\end{figure}

The closest cube in time from \textit{Huygens} landing date is the cube 1481624349\_1 (labeled C14 hereafter), acquired with the High Resolution mode on December 13th, 2004, one month before the landing. We performed the RT analysis on the entire cube. Results for the haze and mist parameters are presented in Fig.~\ref{C14FC=FH}. The histogram scales are the same, for better comparison of the results. They show a distribution of the parameters $F_m$ and F$_h$ close to 1.1 for the case $F_h = F_m$ (Fig.~\ref{C14FC=FH}-a) where the normalized vertical profile is equally multiplied. This result is good considering that we adjust the parameter using a larger wavelength range with VIMS-IR than for DISR, and that we do not have the same incident and emergent angles.  The pixel (10,2) overlaps \textit{Huygens} landing site, where the $\chi^2$ from Eq. \ref{chi2} is 90 for $F_m=F_h = 1.09$, and 75 for $F_h\neq F_m$ with $F_h= 1.08$ and $F_m = 1.24$. Having different haze and mist parameters thus leads to a better fit.

There is no reason for having the same haze and mist parameters, so for the second test case, their values are allowed to be different. The resulting best-fit mean values are 1.22 for F$_m$ (Fig.~\ref{C14FCdiffFH}-b) and 1.09 for $F_h$ (Fig.~\ref{C14FCdiffFH}-c). The mist parameter is more dependent on the lower atmosphere, and is more variable than the haze parameter. This is why the result differs from the first test case. $F_h$ is close to the previous test value while $F_m$ differs, showing that the mist only plays a second order role in the model. Using two parameters instead of one thus results in a gain of information on the mist. The 10 \% differences between \citet{Doose2016} values and ours are acceptable given the different initial data.

Since DISR instrument on board \textit{Huygens} made the unique \textit{in situ} measurement of a Titan's terrain albedo, we propose here a comparison between these observations and surface reflection coefficient obtained using our method on C14 cube. DISR wavelengths range is less extended than VIMS-IR, so the comparison is only possible below 1.6 $\mu$m \citep{schroder_keller_2008, karkoschka2012}.

\begin{figure}[h]
	\begin{center}
		\includegraphics[width=8.5 cm, angle=0]{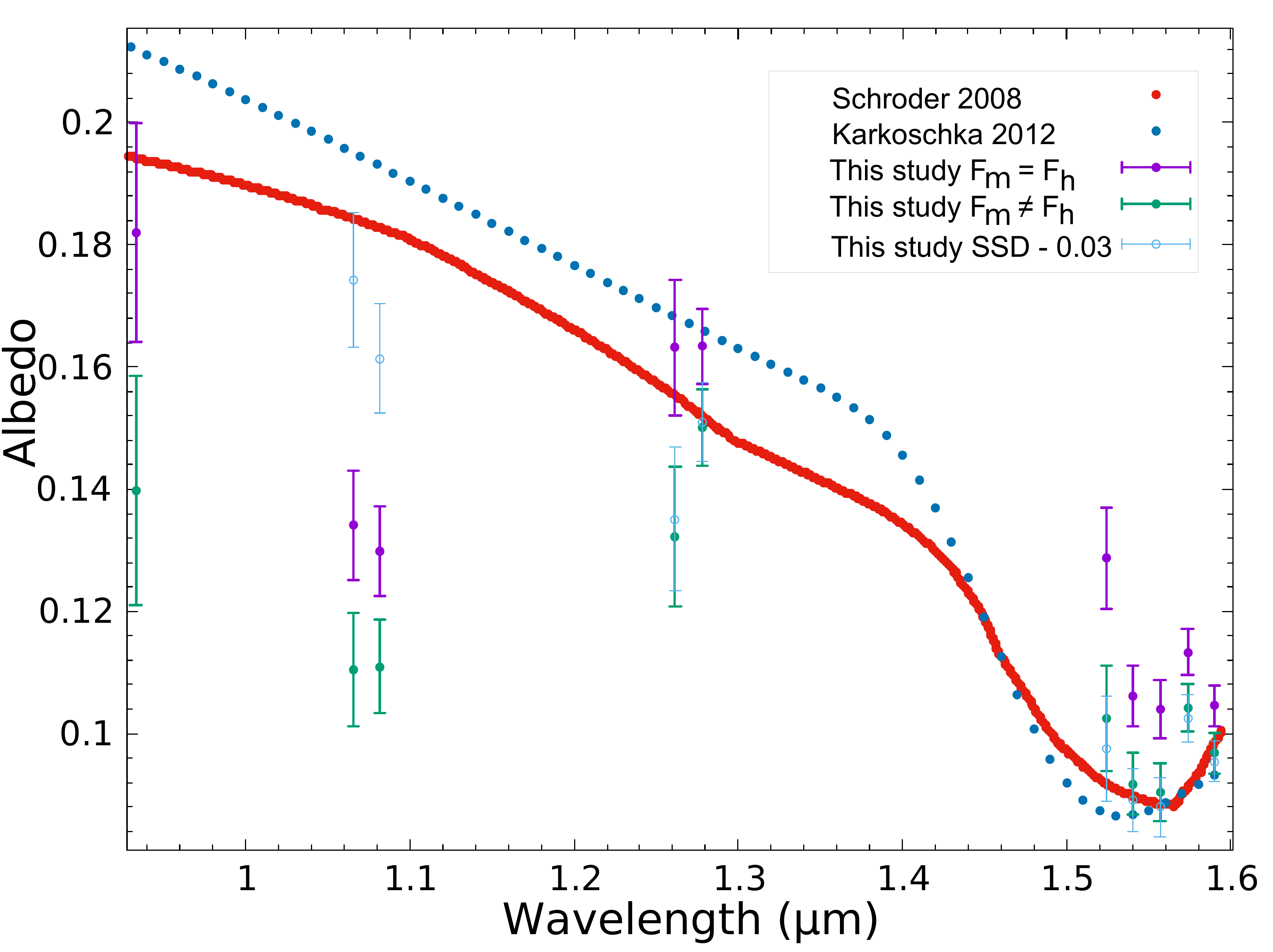}%{fig/ALB_Huygens.pdf}
		
	\end{center}
	\caption[]{	\label{albHuygens}\textit{Huygens} landing site surface albedo, derived from DISR and VIMS instruments. The surface albedo of pixel (10,2) in the cube C14 is retrieved for the cases $F_h=F_m$ (purple), $F_h\neq F_m$ (green) and for a case with the single scattering albedo (SSD) of the mist decreased by -0.03 (light blue). Results from \citet{schroder_keller_2008} (red) and \citet{karkoschka2012} (dark blue), based on DISR dataset are indicated. Literature data are for a null phase angle. We approximately scaled them following \citet{karkoschka2012} for a phase angle of 18$^\circ$.}
\end{figure}

In Fig.~\ref{albHuygens}, we plotted our results derived form VIMS cubes against two DISR observations. We can not compare the absolute albedos directly because of the different conditions like the pixel resolution (14 km for our cube \textit{versus} a few meters at most for DISR after landing), and the geometry of incidence and emergence angles. In our model we assumed a lambertian surface with isotropic reflectivity, but \textit{Huygens} landing site reflectivity is more complex and angle-dependent \citep{karkoschka2012,karkoschka_2016}. We do not expect to obtain exactly the same surface albedo as the previous studies for these reasons. However, the global decreasing shape is similar, except at 1.08 $\mu$m. We also note at 1.55 $\mu$m a similar bowl shape of the albedo curve. The slight difference at 1.08 $\mu$m could mean that this pixel also contains absorbing material around 1.08 $\mu$m. It may also be due to a missing absorption in this window in our model, or a difference in the atmosphere properties, for a small change in the single scattering albedo of the mist has a greater effect at short wavelength ($<1.3 \mu$m) (see the light blue results in Fig.~\ref{albHuygens}). With an observed field much larger for VIMS, the chemical composition and spectra of considered terrains may be heterogenous for fields as large as $\sim 100$~km$^2$; explaining the differences that we have noticed. It could also be caused by the slight VIMS wavelength shift that is more noticeable in the firsts windows since they are much narrower. The differences occurs for other cubes taken in this area and close to the landing date, but not with cubes taken years later, maybe because of the shift. 

Globally, this VIMS/DISR comparison is reassuring even if, for above mentioned reasons, it cannot be considered as	a formal validation of our approach.

\section{Surface albedos of Xanadu and Tui regio}

\subsection{Method}

To retrieve the surface albedo from VIMS pixel data, we fit the observed spectra in the methane bands by adjusting both scaling factor F$_h$ and F$_m$, using the minimization method described in Sect. \ref{error}. Once the scaling factors are adjusted, we retrieve the surface albedo in the methane windows. Fig.~\ref{sd}-a shows the detailed windows. We routinely ran three simulations for each window, assuming a surface albedo of 0.05, 0.15 and 0.25, independently of the wavelength. For each VIMS channel sensitive to the surface, the surface albedo is finally determined by linear interpolation. % We end up with a linear equation of the form $y=ax + b$ :

%\begin{align*}
%a&=\dfrac{I/F_{(A=0.25)}-I/F_{(A=0.05)}}{0.25-0.05}\\
%b&=-aA_{0.15}+\dfrac{I/F_{(A=0.25)}+2I/F_{(A=0.15)}+I/F_{(A=0.05)}}{4}
%\end{align*}
%With $A$ the surface albedo. 

\subsection{Criteria for water ice presence}

\begin{figure}[h]
	\begin{center}
		\includegraphics[width=16 cm, angle=0]{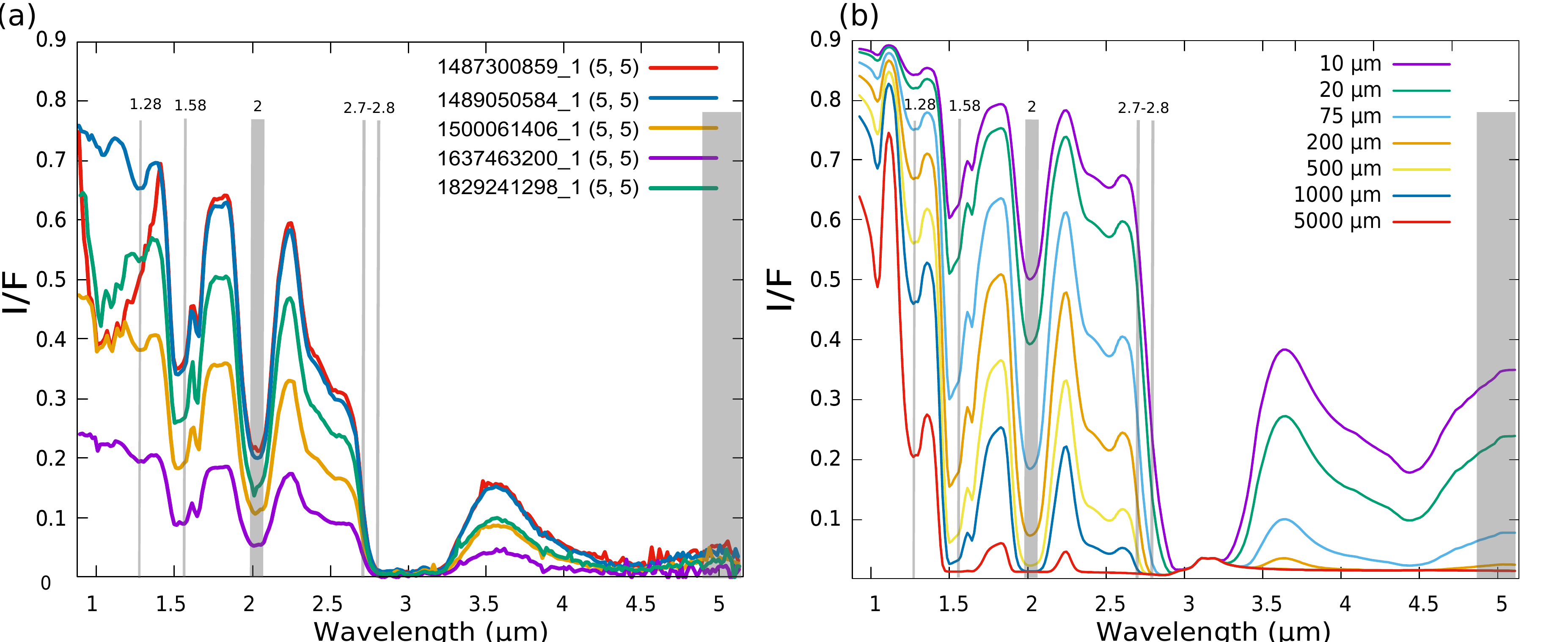}
		
	\end{center}
	\caption[]{	\label{Eau} \label{Encelade}(a) Spectra of Enceladus, from 5 different VIMS cubes, (b) Experimental water-ice spectra with different grain size, from the GHOSST database. Grey columns represent the channels used to calculate the ratios. Their widths depends on the number of channels averaged.}
\end{figure}

Using VIMS spectra collected on Enceladus and shown in Fig.\ref{Eau}-a, we determined the water ice absorptions. They are centered at 1.04, 1.27, 1.51, 2.03 and 3 $\mu$m. Titan's atmospheric windows are not exactly located where these absorptions occur, but they are influenced by them even when water ice absorptions are located at the edges of atmospheric windows.
Due to Titan's atmosphere strong absorptions, and because sedimented aerosols partly cover the surface, we do not expect a pure water-ice spectra on Titan. This is why we looked for traces and influence of the water-ice in Titan's surface spectra.We have then defined our criteria to measure the water ice signature on Titan. These criteria are based on a combination of conditions on the ratios between the water-ice absorption bands \citep{barnes_etal_2007,Robidel2020}. Following the shape of the water ice spectrum, those mean and ratios values likely indicate the presence of water ice signature if:
\begin{itemize}
	%\item  $1.28/1.08$ $> 1$. % pas fiable, car selon les spectres d'encelade on n'a pas la même chose
	\item[\textbullet]$r_1$ :  $1.58/1.28$ $< 1$, as water ice absorbs more at 1.58  $\mu$m
	\item[\textbullet]$r_2$ :  $2.03/1.28$ $<1$, as water ice absorbs more at 2.03  $\mu$m
	\item[\textbullet]$r_3$ :  $2.03/1.58$ $<1$, as water ice absorbs more at 2.03  $\mu$m
	\item[\textbullet]$r_4$ :   $2.7/2.8$ $>1$, as water ice absorbs more at 2.8  $\mu$m
	\item[\textbullet]$\overline{A_5}$ : Mean of the $5$ $\mu$m window low as water ice spectrum is low at $5$ $\mu$m compared to the rest of the spectrum.
\end{itemize}
We chose not to use the ratio $1.28/1.08$ \citep{barnes_etal_2007}, which, according to the considered spectra of Enceladus (Fig.~\ref{Eau}-a), can either be greater or lower than 1 - possibly because of impurities - even if in the pure water ice case from GHOSST\footnote{\url{https://ghosst.osug.fr/}} database  (Fig.~\ref{Eau}-b), the value of the ratio is strictly greater than 1. Not all ratios have the same impact: we tend to have greater error bars at short wavelengths for the albedo, and there are known missing absorptions \citep{Cours2020}, like for ethane at 2.7-2.8 $\mu$m \citep{Maltagliati2015}. Furthermore, we work with the VIMS data calibration RC-19 which is more recent, but less reliable for the 2.7-2.8 windows compared to the RC-17 as it tends to change the intensities values in this window by about 5 \% \citep{clark_etal_2018}. With the assumption that the effect of the missing ethane absorptions and the calibration is the same for each pixel, working in relative and not in absolute is acceptable. Even so, we chose to weight the ratios and mean by their known calculated uncertainties. For the mean albedo $\overline{A_i}$ of a window of $N$ values, the uncertainties $\Delta\overline{A_i}$  is:

\begin{equation}\label{err1}
	\Delta\overline{A_i} = \frac{1}{N}\sqrt{\sigma_{A_{i_1}}^2 + ... +\sigma_{A_{i_N}}^2 }
\end{equation}

with $\sigma_{A_{i}}$ the uncertainty of the surface albedo at a given wavelength (see Eq. \ref{err}). Then for each ratio $ r = \overline{A}_i/\overline{A}_j$, the relative uncertainty of the ratio is:

\begin{equation}\label{err2}
	\sigma_r = \dfrac{\Delta r}{r}=\dfrac{\Delta\overline{A_i}}{\overline{A_i}} +\dfrac{ \Delta\overline{A_j}}{\overline{A_j}}
\end{equation}

We defined a multi-criteria analysis parameter $\delta$ to estimate the presence of water ice. This parameter combines the relevant individual $I/F$ ratios and the 5$\mu$m mean previously mentioned. For the multi-criteria analysis, these ratios and 5$\mu$m-mean are normalized between 0 and 1 with the following method.  For a ratio $r_1$, the normalized ratio $r_{1_{n}}$ is:
\begin{equation}
	r_{1_{n}}=\frac{r_{1}-r_{1_{min}}}{r_{1_{max}}-r_{1_{min}}}
\end{equation}

With $r_{1_{min}}$ and $r_{1_{max}}$ the minimum and maximum values of the ratio for the considered cube.
The normalized values are then added if the expected value of the non-normalized ratio or mean in case of pure water ice signature is greater than 1 (i.e. $r_{2.7/2.8}$). They are subtracted if the expected value is lower than 1 (i.e. all other ratios and 5$\mu$m-mean), as shown in equation \ref{multi}. In this way, theoretically the highest possible value for $\delta$ is 1 indicating a prominent water ice-rich signature, and the lowest value is $-4$, for an extremely poor water-ice signal. In addition, the normalized ratios and mean at 5 $\mu$m are weighted by (1- $\sigma_r$), with $\sigma_r$ the relative error of each VIMS channels ratio $r$. We also set a lower limit for (1- $\sigma_r$) which cannot be lower than 0, even if it does not happen in our results. The parameter $\delta$ is explicitly:
\begin{equation}\label{multi}
	\delta=   - (1-\sigma_{A_5})\overline{A_{5_{n}}}	+\sum_{k=1}^4\pm(1-\sigma_{r_k})r_{n_k}
	%(1-\sigma_r)(-\frac{\overline{A}_{1.58}}{\overline{A}_{1.28}} - \frac{\overline{A}_{2.03}}{\overline{A}_{1.28}} - \frac{\overline{A}_{2.03}}{\overline{A}_{1.58}} + \frac{\overline{A}_{2.7}}{\overline{A}_{2.8}} -\overline{A}_{5.})
\end{equation}
%For the multi-criteria analysis, each ratio taken into account is reduced to the interval [0, 1] on the set of pixels considered. Then we add up the results for each pixel, by multiplying by −1 if the quantity thus standardized must be small, leaving it positive if it must be large in the case of the presence of water ice. With this kind of definition, the parameter obtained is all the greater when the presence of water is likely.
With $\overline{A_{5_n}}$ the normalized mean surface albedo at 5 $\mu$m, and $k$ the number of the normalized ratio $r_n$. $\delta$ increases when the probability of water ice presence increases. With Eq. \ref{err1} and \ref{err2}, we are able to establish our ratios' error, and determine which results are relevant. With ratios weighted by their uncertainties, -4 and 1 are ideal values that can not be obtained. Furthermore, $\delta$ must not be interpreted alone since it is obtained using normalized values: there is no specific threshold value of $\delta$ which define is indeed water ice or not. It shows subtle changes in the surface composition. The possible results consistency between the cubes reinforces the interpretation. We also need VIMS colormaps and the RADAR map to scout the terrains beforehand and to better interpret the results. The criteria $\delta$ should be considered as an indicator of the strength of the water ice signature and not as a direct estimator of water ice abundance. It should be regarded as a tool for terrain classification.

\subsection{Haze and mist results}
\begin{figure}[h]
	\begin{center}
		\includegraphics[width=16cm]{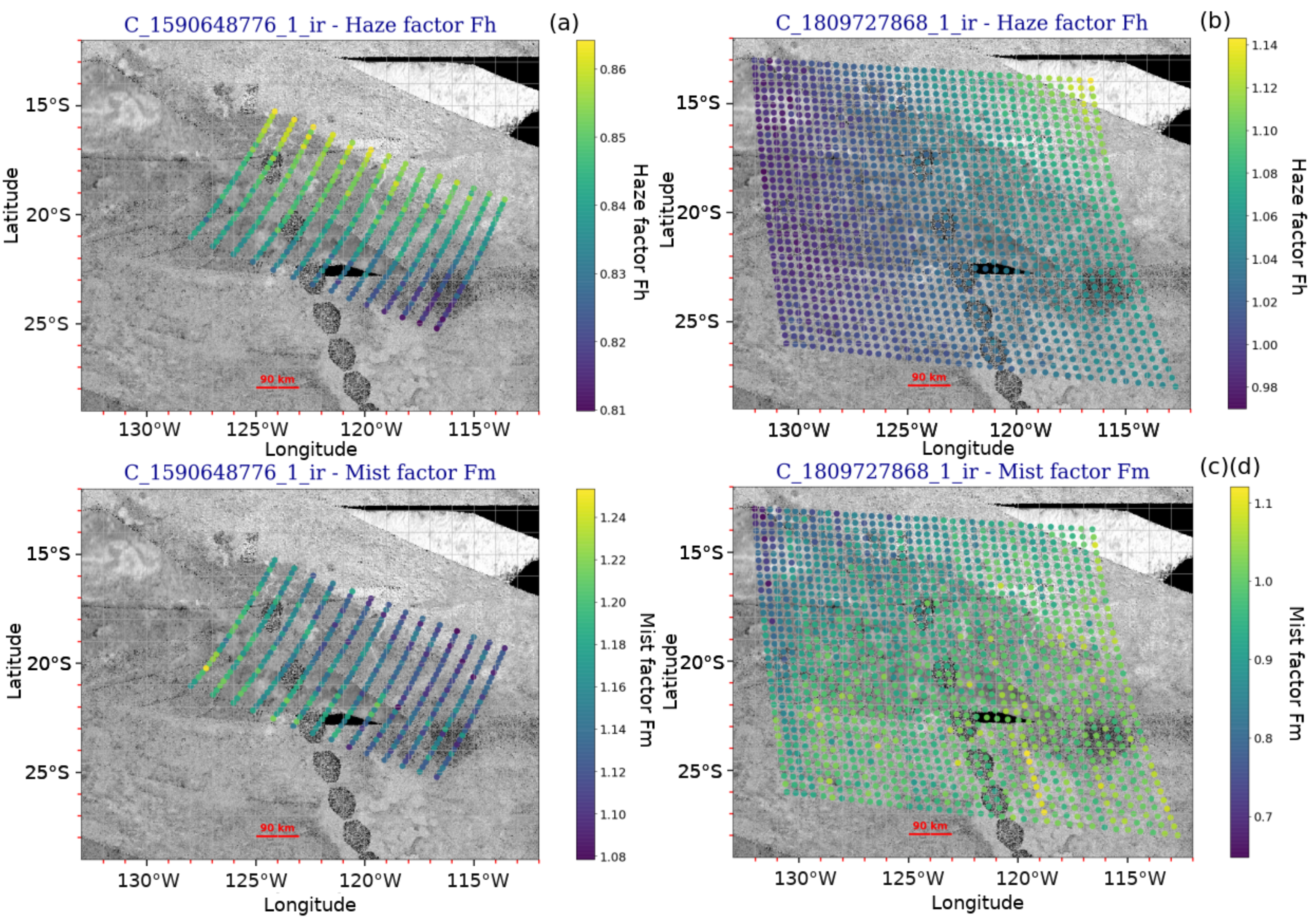}
		
	\end{center}
	\caption[]{	\label{FhFm}Maps of the haze and mist coefficients for C15 (a) and (c) respectively, and for C18 (b) and (d) respectively, overlaid to the RADAR map. }
\end{figure}
\begin{figure}[h]
	\begin{center}
		\includegraphics[width=16 cm, angle=0]{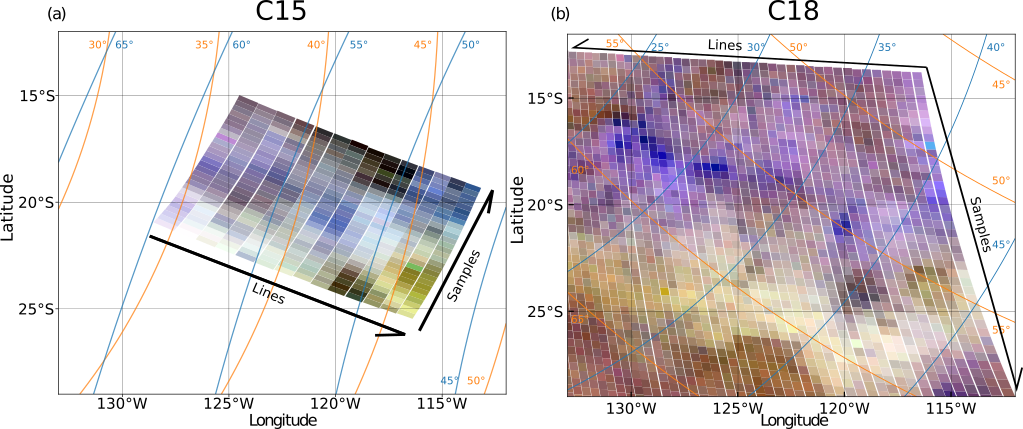}
	\end{center}
	\caption[]{\label{ene_incid}(a) Pixel footprints of the VIMS cube \texttt{1590648776\_1}. Colors are mapped with VIMS spectral channels ratios: $1.58/1.28$ as red, $2.03/1.28$ as blue, $1.28/1.08$  as green, with contourlines corresponding to the incident angles (orange lines) and emergence angles (blue lines). (b) Similar figure for the VIMS cube \texttt{1809727868\_1}. This color code highlights water ice in dark blue.}
\end{figure}

Fig.~\ref{FhFm} shows F$_h$ and F$_m$ regional variation in C15 and C18, overlaid on the RADAR map. Each point corresponds to a cube pixel.  The haze factor seems to be more stable than the mist factor, which is expected since the mist factor should be more sensitive to local atmospheric variations on the surface. Both factors are not correlated, but we note a correlation with the incident and emergent angles in Fig \ref{ene_incid}. C15 coefficient F$_h$ (Fig.~\ref{FhFm}-a) is uncorrelated, but F$_m$ in C18 (Fig.~ \ref{FhFm}-d) seems to be correlated with the emergent angles, while F$_m$ in C15 and F$_h$ in C18 (Fig.~\ref{FhFm} c and b) appear to be anti-correlated with the incident angles. Yet, as no regular pattern stands out, there is no argument for an actual correlation between the angles and the haze and mist parameters values.

We acknowledge a \textit{bias} in our model: the atmosphere is assumed to be laterally homogeneous and infinite with no change of the aerosols vertical profile, whereas in our results the parameters F$_h$ and F$_m$ change for each pixels. However the \textit{bias} is small, as at most rays go through 5 different columns of atmosphere before reaching the ground pixel for C15, and 8 for C18, and the values of the haze and mist parameters do not change in a significant manner within 5 or 8 adjacent pixels.  The model can be tuned with free parameters but the haze and mist vertical profiles, the particle scattering phase functions and the gas vertical profiles that we used may be different from the actual profiles on Titan. It would modify the angular description of the modeled intensities and of the retrieved surface albedo. Therefore, it could induce another bias because the two images are not taken with the same observation geometry.% Furthermore, our criteria for the water-ice signature is relative and non sensitive to those \textit{bias}. Fig.~\ref{FhFm}, \ref{ratio1} and \ref{ratio2} do not shows those \textit{bias} so we can consider valid the values of  F$_h$ and F$_m$.

\subsection{Water-ice distribution results} 
\begin{figure}[h]
	\begin{center}
		\includegraphics[width=16cm]{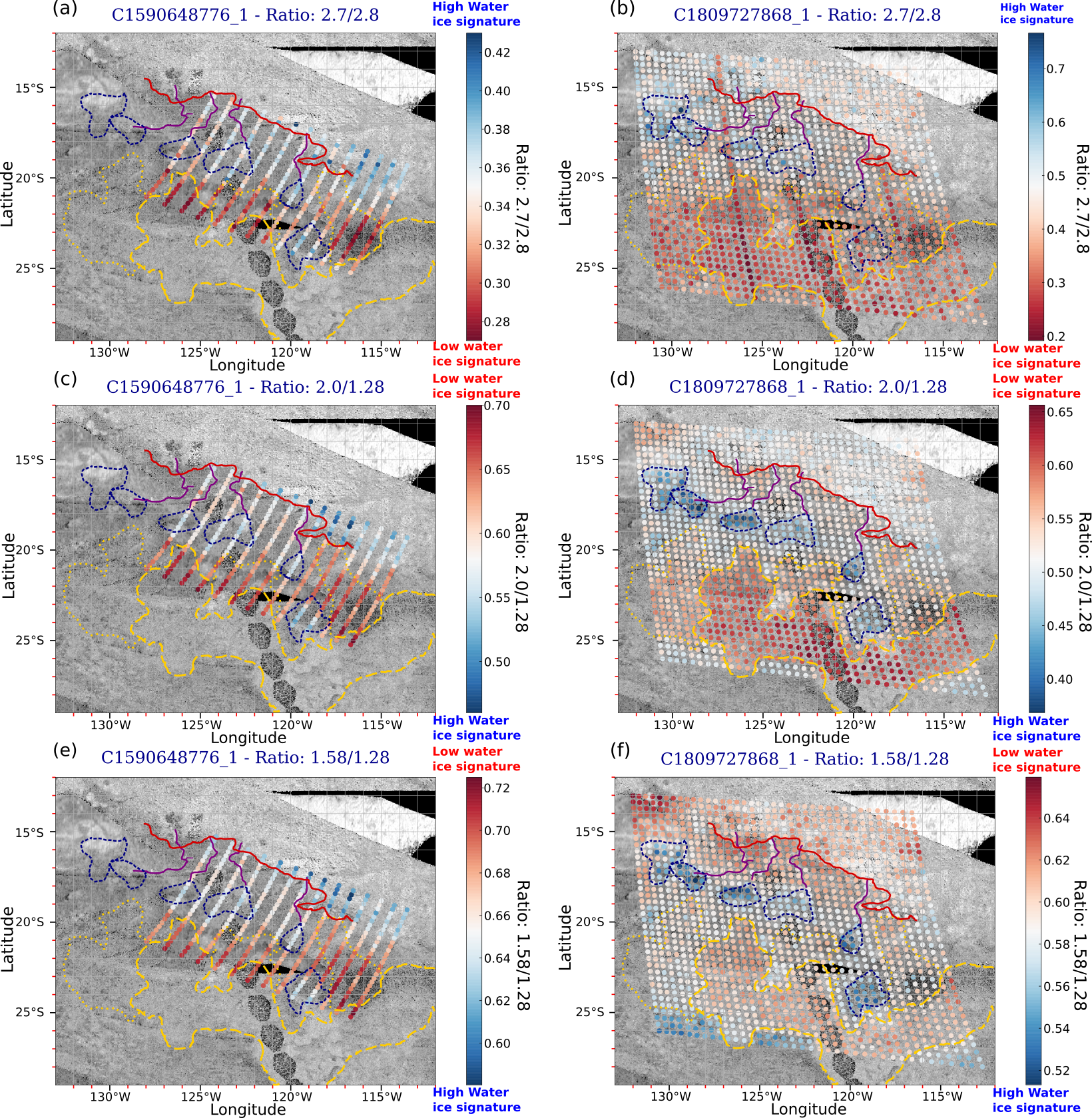}
		
	\end{center}
	\caption[]{	\label{ratio1}Values of the ratios 2.7/2.8 (top panels), 2.0/1.28 (middle panels), and 1.59/1.28 (bottom panels) for the VIMS cubes C15 (a,c,e), and C18 (b,d,f) respectively, superimposed on the RADAR map. Each point corresponds to the position of a pixel of the cubes. Channels are outlined by purple lines, and areas of low and high water-ice signal are circled in yellow and blue, respectively. The absolute error for each ratio is: (a) $\pm0.05$, (b) $\pm 0.15$, (c) $\pm0.05$, (d) $\pm0.05$, (e) $\pm0.07$, (f)$\pm0.06$.}
\end{figure}

\begin{figure}[h]
	\begin{center}
		\includegraphics[width=16cm]{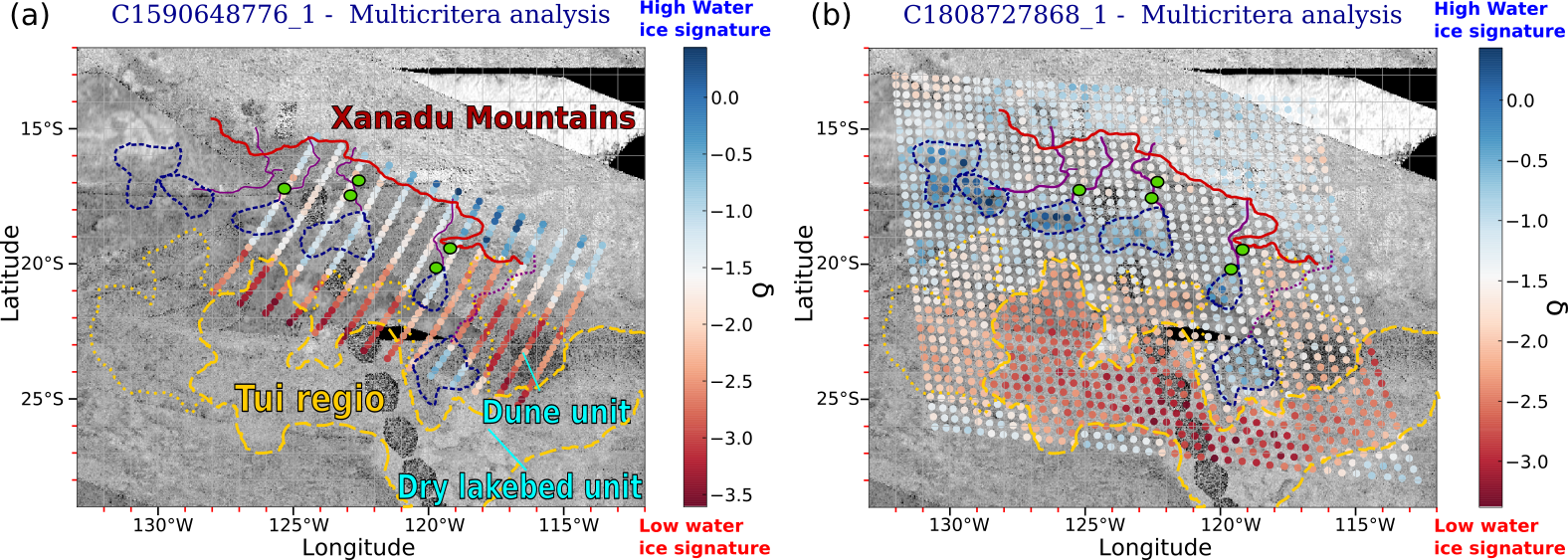}
		
	\end{center}
	\caption[]{	\label{ratio2}%Values of the mean at 5 $\mu$m (top panels), and contour map derived from the multi-criteria analysis parameter $\delta$ (bottom panels) for the VIMS cubes C15 (a,c), and C18 (b,d) respectively, superimposed on the RADAR map.  Each point corresponds to the position of a pixel from the cubes (top). Channels are outlined by purple lines. Regions of high and low $\delta$ are annotated in blue and yellow lines respectively.
		Values of the multi-criteria analysis parameter $\delta$ for the VIMS cubes C15 (a), and C18 (b) respectively, superimposed to the RADAR map.  Each point corresponds to the position of a pixel from the cubes. Channels are outlined by purple lines. Regions of high and low $\delta$ are annotated in blue and yellow lines respectively, and fans highlighted by \citet{radebaugh2018} are represented by green dots.}
\end{figure}

The maps of different non-normalized ratio values for both VIMS cubes are in Fig.~\ref{ratio1}, and the $\delta$ values are presented in Fig.~\ref{ratio2}. Given our estimated uncertainties, the extreme values of the ratios (in red and blue) are significant. Furthermore, similar structures stand out on both cubes, although we cannot compare the absolute values as the illumination and viewing geometries are different between C15 and C18 that, in addition, were acquired 7 years apart.
It is also important to remind that the ice crust on Titan is covered by sedimented aerosols \citep{Janssen2016,brossier2018}. The sediment layer thickness has an influence on VIMS spectral channels ratios \citep{rannou_etal_2016}. %, where the water ice signature is more or less important due to the surface composition (dust, aerosols, liquids, and ice). 
In both Fig.~\ref{ratio1} and \ref{ratio2}, units exhibiting different strength of the water ice signal according to the multi-criteria analysis of C18 (Fig.~\ref{ratio2}-d) are outlined.
In C15, we do observe the highest water ice signature on Xanadu mountains, and a medium-high signature at 24$^{\circ}$S 120$^{\circ}$W, next to the dune unit. Surprisingly, in C18, the strongest signals are not in the same place as in C15. They are at the ends of the channels in blue dotted areas, which only displays a medium water ice signature in C15.
There is a clear correlation between the structures on the RADAR map, and the signature of water ice. For example the blue dotted areas tend to correspond to dark and smooth gray structures in the lowest resolution part of the RADAR map.
The area with low water-ice signature around 23.5$^{\circ}$S 116$^{\circ}$W corresponds to the dune unit in the RADAR map, and its surrounding dark gray halo. The halo has a lower water-ice signature than its center.  A dark gray terrain at 22$^{\circ}$S 122$^{\circ}$W on the RADAR map is also well delimited by a low $\delta$ on C15, and a medium-low one in C18. This is also the case for the medium-low $\delta$ on the top corners of C18. The largest terrain with low $\delta$ is located in Tui Regio.
Ratios are of medium value in the channel beds, suggesting a slightly higher water ice signature than in the surrounding terrains (see Fig.~\ref{ratio1}-d-e for the most flagrant). Even if the error on these regions reduces the significance of these variations, we do observe the same trend in both cubes. This support the hypothesis that the channels descend from the mountains, and contain water ice cobbles, probably covered by organic sediments. Above all, we notice the presence of an area of high $\delta$ at the end of each channel. 

%The spatial distribution of the water-ice signature intensity is noticeable in Fig.~\ref{cube}. We can associate colors of the colored composition to the results of our ratio. The bright area corresponds to a low signal of water ice. The purple one correspond to a medium low signal. The green area has a medium high signal of water ice, and the dark green, or black area has a high water ice signature.
%      \begin{figure}[h]
%	\begin{center}
%		\includegraphics[width=8.5cm]{fig/cubes.pdf}

%	\end{center}
%	\caption[]{	\label{cube}Cube v1590648776 (left) and v1809727868 (right) georeferenced, with a color composition R=2.03, B=1.58, V=2.79}
% \end{figure}
The color code in Fig.~\ref{ene_incid} is set to highlight in dark blue terrains with high water ice signature. However, our results (Fig.~\ref{ratio1} and \ref{ratio2}) reveal areas with strong water ice signal in dark brown for C15 (Fig.~\ref{ene_incid}-a). For C18 the highest water ice signatures are in dark brown, and dark blue areas. This confirms the need of a radiative transfer analysis to extract the surface signal from the atmospheric signal, and to characterize the nuances of the water-ice signal intensity.

\begin{figure}[h]
	\begin{center}
		\includegraphics[width=16cm]{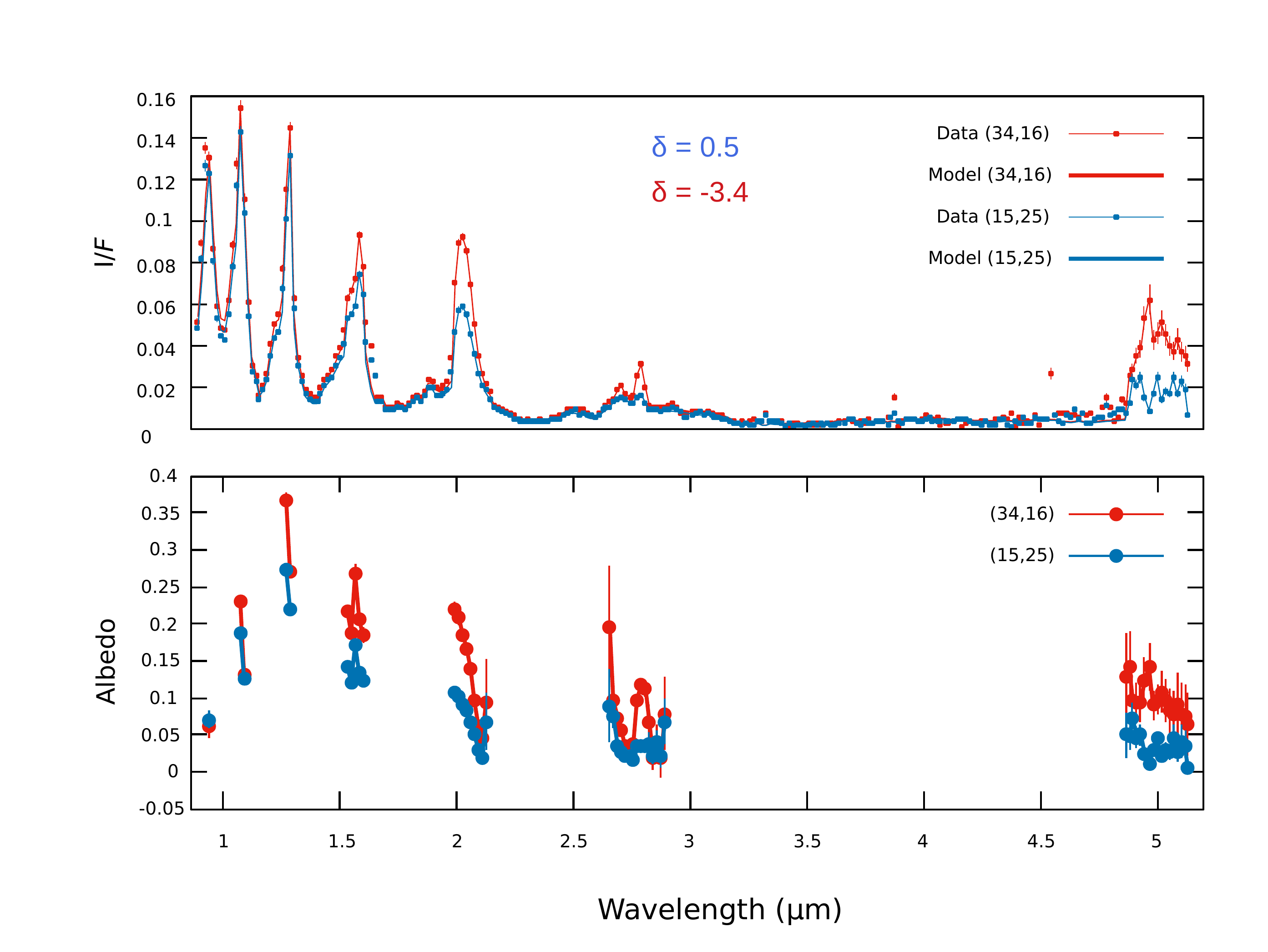}
		
	\end{center}
	\caption[]{	\label{plot} $I/F$ spectra of pixels (34,16) and (15,25) of the cube C18 with low (red) and high (blue) values of our water-ice signal criteria $\delta$ (top panel), and their respective albedo (bottom panel).}
\end{figure}

Fig.~\ref{plot} displays the spectra and surface albedo of two pixels taken in zones of low and high water-ice signature for the cube C18 (respectively, $\delta  = 0.5$, and  $\delta= -3.4$). The presence of water-ice darkens the albedo in most of Titan's atmospheric windows (see Fig.~\ref{Eau} water-ice absorptions), consequently lowering the $I/F$, whereas where the water-ice signal is low, the albedo is increased. This effect is especially significant at 1.5, 2, 2.8 and 5 $\mu$m. However the shape of the albedo for the 2.8 $\mu$m window changes the most.
%%%%%%%%%%%%%%%%%%%%%%%%%%%%%%%%%%%%%%%%%%%%%%%%%%%%%%%%%%%%%%%%%%%%%%%%%%%%%%%%%%%%%%%%%%%%%%%%%%%%%%%%%%%%%%%%%%%%%%%%%%%%%%%%%%%%
%% Discussion et Conclusion
%\section{Interpretations and Discussion}

%{\color{DarkRed2}Point de discussion sur la taille des grains de glace : les lits de rivieres sont-ils favorises ? \\
%\citet{taffin_etal_2012} on fait une discussion de ce type la avec les ``Tiger Trips'' d'Encelde.\\
%}

%The ratio $2.7/2.8$ $\mu$m is associated with a change of the grain size of water-ice \citep{barnes_etal_2007}
\section{Discussion on the water ice signature and on geological structures} 

With the color maps, we can discuss the meaning of the ice index and the connection with surface features and processes. At first glance, the two dataset C15 and C18 yield, for each color ratios and for the multicriteria ($\delta$), similar and consistent maps. However, we note that the absolute values reported on the maps systematically differ between C15 and C18. We also remark that the nature of some areas clearly differs, which reveals changes between the two flybys. We describe in this section the factors that help understanding our maps and the parameters that could influence and modulate the retrieved surface albedos, ratios and multicriteria index in the maps.

\subsection{Grain size and sediment cover}

First of all, the intensity of water ice signatures depends on the size of ice grains. Data displayed in Fig.~\ref{Eau}-b clearly illustrates this dependence. This stresses that a change in the strength of water ice signature could be due to a change in the average size of ice grains composing the uppermost layer of Titan's ground, or to a thicker sedimentary layer covering or mixed	with the ice layer \citep{brossier2018}. Based on this figure, the ratio $2.7/2.8$ could be (with caution) used as an indicator of the grain size due to the strong absorption of water ice at 2.9 to 3.25 $\mu$m. The higher the ratio is, the smaller the grain size is. It could be directly used to classify different terrains if the sedimentary layer is homogeneous. If the sedimentary layer is not homogeneous, disentangling the two effects is not possible because the effect of a coverage by sediments is not known. A specific model is needed to interpret the details of the color ratio in term of physical parameters, like what was done in \citet{brossier2018}, where they noted in the IR-blue terrains a grain size generally larger than 100 $\mu$m, and half as many solutions with a grain size smaller than 0.1 $\mu$m. Another difficulty arises because the 2.7-2.8 $\mu$m window is incompletely treated with ethane and possibly other molecules absorptions missing in the model.  \\

Therefore, in this work, we can compare and classify the different terrains with the color ratio and the multicriteria index. It means that we discuss the apparent presence of water ice, but we have to keep in mind that more or less prominent ice signature may have multiple origin and can not be directly linked to the real amount of ice at the surface.\\
%The intensity of the water ice signature depends on the grain size of the ice. The data displayed in Fig.~\ref{Eau}-b recall clearly this dependence. With this figure, we can infer with caution that the ratio $2.7/2.78$ is an indicators of the grainsize if the dust layer is homogeneous everywhere. When this ratio is high, it could indicate a smaller grainsize.  The 2.7-2.8 $\mu$m window is incompletely treated in the RT analysis, but it can still help classify different terrains.
Using a ratio instead of absolute band values allow us to still be more sensitive to water ice absorption. Fig.~\ref{ratio1}-a and -b show different patterns between C18 and C15%. C15 presents a high ratio at the top right corner, on mountains structures, whereas in C18 the same region is associated with a moderate value of the ratio compared to what is found elsewhere, and the zones of high ratio are located at the blue dotted areas around 17$^{\circ}$S and 19$^{\circ}$S. 
. Similarity still arises: on both cubes, at the beginning of the channels, the $2.7/2.78$ ratio is lower than at the bottom. However the differences are within the estimated error bars. Taking this into account, as noted previously, there is either a variation of the grain size in the cube indicating a larger grain size in Xanadu, a variation of abundance of water ice, and/or a disparity in the dust layer thickness. In the case of sediment transport, smaller and lighter grains travel generally further as less energy is required to move the particles \citep{Sediment_Transport}. We then expect larger grains on Xanadu at the channels origin, and that further along the river the grain size decreases, as inferred here. 
%%We have an anti-correlation between those two ratio in Fig.~\ref{ratio1}-a-b and \ref{ratio2}a-b, for each cube. However for C18, the variation of the $2.7/2.8$ doesn't have the same shape as the other ratios. We suppose there is a mix between the intensity of the ice signature, and the grainsize, resulting in a ratio of medium value at Xanadu, instead of a high one because of the abundance of water-ice. 
%Ratio 2.7/2.8 is of medium value in the plains and Xanadu, whereas the 5 $\mu$m mean is really low in Xanadu mountains, and of medium low values in the plains.

\subsection{Seasonal cycle and dust storm}

As noted previously, we observe differences between C15 and C18, either for the haze and mist scale factors, or for $\delta$.
Surface albedo are retrieved with a specific setup of absorption and scattering properties in the atmosphere, given in detail in previous sections. But, this should be considered as an operational description, with some parameters determined by previous studies, predicted by models or set by free parameters. Constraining the atmosphere properties for retrieving the same surface albedo with C15 and C18 would be possible. However, the atmosphere characteristics (particle vertical and gas profiles, phase functions,...) may also change with time. The two observations were taken 7 years apart, over the north spring equinox. Some atmospheric properties probably have changed between these two dates, especially the haze and mist vertical profiles \citep{vinatier_etal_2010a, Seignovert2020}. This explains well why the amount of particles needed to fit data changes between these two dates. But, the actual change in haze and mist profiles can not be accounted in detail by simple parameters such as the haze and mist factors ($F_h$ and $F_m$). The global shifts in color ratio and in the multicriteria index between the two observation sets can then be also explain by a real change in atmosphere properties.\\

The maps derived from observations also differs in some areas with obvious changes in water ice signatures compared to the surrounding. For instance, a strip of several pixels width at the northern edge of the maps retrieved with C15 systematically has a stronger ice signature than the pixels a bit south. This strip does not exist in C18. Notably, this strip corresponds to the southern edge of Xanadu mountains, delineated in Figure \ref{radarVIMS} and \ref{ratio2}. These changes in color ratios and in $\delta$ indicate changes in the composition of the uppermost surface layer.\\

Two specific processes, which are not mutually exclusive, may explain these regional changes. First, the amount of deposited material rate of the aerosols is low on Titan, but dust storms probably occurred in between 2008 and 2010 due to the equinox \citep{Charnay2015, Rodriguez2018}, and one particularly at the East side of Xanadu. It could have influenced the amount of deposited material in this period, and the number of large particles in the lower atmosphere issued from surface erosion. Such an event may have deposited a sufficiently thick layer of particles to change the layer's color relative to the surrounding. Secondly, \citet{turtle_eal_2011a} report an episode of heavy rain during the equinoctial period in 2010, in the equatorial region. This occurred at 20$^o$S, but at an another longitude than our region of interest. The rain flooded a wide region and produced an obvious and persistent change in surface reflectivity. A similar or less intense flooding may have occurred elsewhere in the same period of time. \citet{Rodriguez2011} report a significant cloud coverage in the south tropical band between 2007 and 2010. They report a cloud event in 2008 above 120$^o$ W and 20$^o$S, comparable in term of apparent area to the event reported by \citet{turtle_eal_2011a}. Such an event can probably alter or remove light sedimentary layers, eventually ice gravels or small pebbles and change locally both surface reflectivity and color ratios. Our maps, indeed, clearly show that an event changed the surface characteristics of some area compared to their surrounding.

%\section{Discussion on the geological structures}

\subsection{Dunes}
The water-ice signature variations can help identifying geological structures, by the shapes and values of those variations.
Based on morphological arguments, \citet{Lopes2019} classified the terrains in C15 and C18 as "plains", "hummocky" for Xanadu mountains, "dunes" for the yellowish structure (at the bottom of Fig.~\ref{ene_incid}-a and annotated in Fig.~ \ref{ratio2}-a),  and "dry lakebed" \citep{Moore2010} for the RADAR-gray structure at the bottom left of the dunes (around 25$^{\circ}$S,118.5$^{\circ}$W) located in Tui Regio. Hummocky terrains are expected to be areas where the icy crust of Titan is more exposed than elsewhere on Titan (i.e., covered by a thinner layer of organic material) \citep{Janssen2016}, which is supported by our results on Xanadu.  The so-called "dune unit" (Fig.~\ref{ratio2}-a), on the contrary, should be poor in water ice. Yet, both the albedo (especially at 5 $\mu$m) and inferred delta parameter in this area suggest the presence of a water ice component. The SAR image does not resolve any individual dunes but there are most likely present at a scale smaller than the RADAR resolution here and with interdune corridors enriched in water ice \citep{barnes2008,Bonnefoy2016}.

%\subsection{Clouds}
\subsection{Precipitations and Alluvial fans}
%We found 2 ISS images of our region of study, taken at 0.935 $\mu$m. A bright thin line appears between the first image on 2006, and the second one in December 2008. We also found a VIMS image from the flyby T44 in our region of interest, but with a lower surface resolution. This image contains longitudinal clouds. When we compare 2 pixels, on the cloud, an one next to it, we observe an augmentation of $I/F$ in all atmospheric windows with the cloud-pixel, including the one at 0.935 $\mu$m. It could be an argument in favor of a cloud for the bright line in ISS image. However, on VIMS image, the cloud is neither thin nor continue, and their orientation are not the same. We excluded the possibility of an artifact, as the same line appears on another ISS image taken minutes later. This line appears on the high resolution ISS global map\footnote{\url{https://astrogeology.usgs.gov/search/map/Titan/Cassini/Global-Mosaic/Titan_ISS_P19658_Mosaic_Global_4km}}. Other bright thin lines in a W-E orientation are noticeable on the global map. 
We find an accumulation of high water-ice signature at the end of the channels. With arguments based on radar-brightness, \cite{legall_etal_2010} have suggested that the bed of these channels could be ice-rich. Unfortunately, they are too narrow to be characterized in detail at VIMS resolution. \citet{brossier2018} interpret 'IR-blue' VIMS patches like our region of interest as possible accumulated erosional products from the icebed, that cover the layer of organic materials, or eroded icy terrains where rainfall washed out the organic layer. The revealed ice-rich signature features could be alluvial fans where sediments have been transported and deposited from the mountains of Xanadu. Such alluvial fan results from the abrupt decrease of the terrain's slope, inducing a slower stream flow and sediments deposit in the plains. There are generally triangular-shaped, similar to the features discussed here, with coarse sediments close to the mouth of the stream and smaller ones further away. The channels that carry materials to the fans can be dry most of the time, as this is most likely the case here and in desert regions on Earth, and occasionally be fed occasionally by violent rainfalls or storms.

As discussed in a previous section, rain showers and cloud activity occur during equinox at these latitudes \citep{turtle_eal_2011a, Rodriguez2011, turtle2018}. As further evidences, other VIMS cubes with lower pixel resolution from the flyby T44 show a large and discontinuous cloud, with an orientation W-E. Strikingly, \cite{griffith_etal_2005} detected a rapid evolution of mid-latitude clouds, that eventually dissipated as a rain fall within a few hours. Such hydrologic events may be the source of liquid alimenting sporadically the channels considered here. A high discharge of liquid and a non-porous soil is needed to transport sediments over 300 km in a dry region. This aspect is also discussed by \citet{Faulk2017} and  \citet{Birch2016}, who have considered occasional intense rainfalls at 20$^{\circ}$S of latitude, that can create alluvial fans.

The extrapolated topography from \citet{corlies_etal_2017} does not show abrupt slope decrease at this particular site (maybe because of a low resolution), but \citet{Birch2016} do find equatorial fans in Xanadu region. However they do not classify these blue dotted areas as fans. \citet{radebaugh2018} mention fans in our area of study (green dots in Fig.~\ref{radarVIMS} and \ref{ratio2} ), but these patterns appear to be much smaller than what we observe. Although, when we look closely at our results in Fig.~\ref{ratio2}-b, we notice that in most cases the green dots' signature of water-ice is a bit higher than the area surrounding the dots. Another origin may also explains the presence of these enhanced water-ice signature at the foot of channels, between 17 and 19$^{\circ}$S.

% With the orographic effect, we could have an higher precipitation rate at the foot of Xanadu.

% An other potential alluvial fan at 30$^{\circ}$S114$^{\circ}$W shows on C18, around pixel (40,6) in Fig.\ref{ene_incid}. It is also a better candidate, as it directly exits a mountainous region.
\subsection{Impact craters}
To investigate the nature of the spots of high water-ice signature, it may be relevant to put our zone of study in a broader context. Few craters impact exist on Titan, compared to other moons \citep{Wood2010, Neish2013}. This feature can be explained by the presence of a dense atmosphere, and can be the consequence of and deposition by eolian and fluvial activities. Resurfacing by tectonic or cryovolcanism may also play a role. Impact craters on Titan, particularly if they are young, are windows to the crust and therefore have ice-rich inner and outer rim \citep{Janssen2016,griffith_etal_2019}. Compared with the rest of Titan, their rims are radar-bright with low emissivity at 2.2cm, like that of the hummock terrains \citep{Lopes2019}.  On Titan, low emissivity appears to be associated with water-ice \citep{malaska2016,Janssen2016}.

\begin{figure}[h]
	\begin{center}
		\includegraphics[width=16cm]{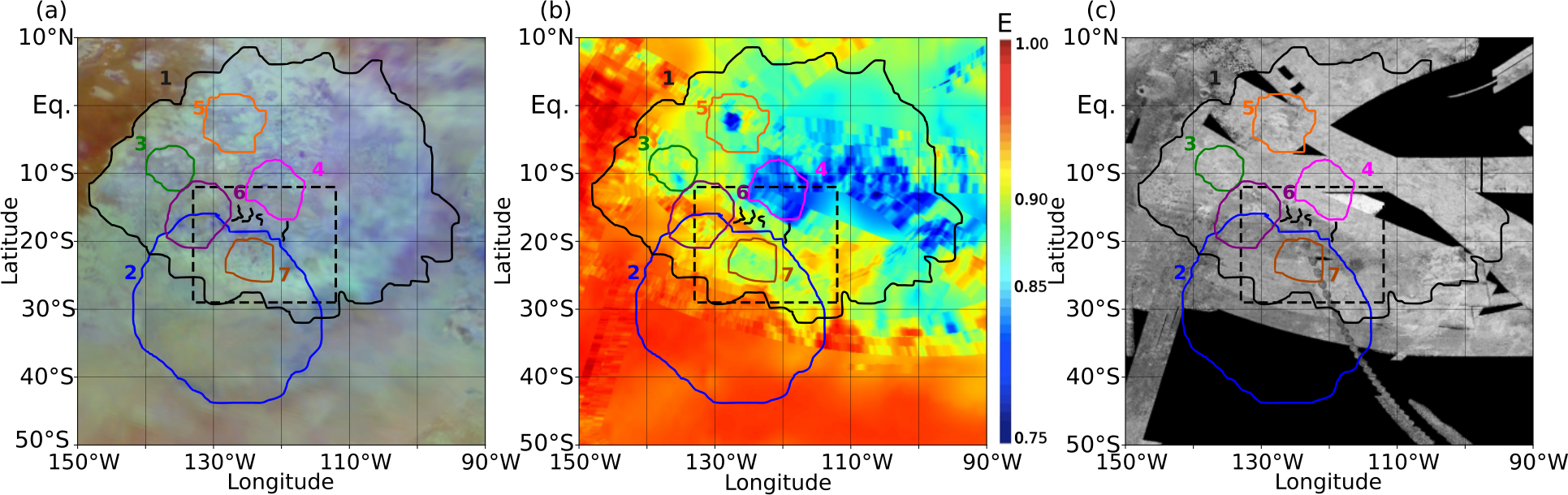}
		
	\end{center}
	\caption[]{\label{crat} (a) Zoom of Titan's ISS\_VIMS map from \citet{TitanISSVIMS} on Xanadu region,(b) Titan's 2.2cm wavelength emissivity map from \citet{Janssen2016}, and (c) Titan RADAR map, with outlined potential crater areas. The dashed rectangle is our zone of study, and the thin black line represent the channels.}
\end{figure}

Xanadu is a complex zone with multiple ancient impact crater candidates \citep{Wood2010}. The global mosaic map of VIMS-ISS \citep{TitanISSVIMS} in Fig.~\ref{crat}-a, reveals 2 giant circular shapes on Xanadu (crater \#1 and  \#2 in this Fig.~\ref{crat}). These features could be artifacts due to the enhanced resolution of VIMS dataset in these areas compared to the rest of Xanadu. However, crater \#1 is also visible on the emissivity map (Fig.~\ref{crat}-b) as a large-scale low emissivity region. \citet{Brown2011} argue in favor of a giant crater called XCF. The spots of enhanced water-ice downstream are located at the limit of a circular feature that could correspond to the rim of crater \#2 but there is no noticeable ring-like feature on the emissivity map, probably due to the low resolution of the bottom half part of the map for this potential crater. Instead, they could also correspond to the inner ring of crater \#1. Finally, crater \#4 may be the inner ring of crater \#1.

Though, giant impacts are not required to create these putative rims. Smaller potential craters exist in Xanadu. They are numbered form 3 to 7 in our region of interest. We select potential craters when they check at least 2 of these following criteria: a circular shape in Fig.~\ref{crat}-c, a low emissivity spot on  Fig.~\ref{crat}-b and a corresponding geological crater-like feature in Fig.~\ref{crat}-a. Putative craters \#6 and \#7 may be associated with water-ice enhanced signal on the rims. Putative craters \#3, \#4 and \#5 are too far from the area of enhanced water ice signal, but it would be interesting to investigate those features in more details. Remarkably, high $\delta$ spots and channels often correspond to value of medium emissivity.

Despite these correlations, we do not find them convincing enough to ascertain that impact craters were the cause of the enhanced water-ice signatures at the channels' ends. %There are also other low-emissivity spots needing further studies in Fig \ref{crat}-b that correspond to geological structures.

\subsection{Delta and evaporites}

\citet{MacKenzie2016} classify Tui Regio as an evaporitic terrain. This is in good agreement with our results showing a very low water-ice signature inside this area (i.e. the yellow dashed area in Fig.~\ref{ratio2}). This region is suspected to be a paleo sea \citep{Moore2010,hofgartner_etal_2020}. While it is not incompatible with an old impact crater formation, as seas may form in craters, it casts a doubt on the origin of the water-ice signature enhancement at the border of Tui Regio, where the channels end. If the channels where active at the same time of the sea, these features could be delta-like formations, occurring when the stream is slowed down at the contact of the sea, resulting in an accumulation of fine deposits at the margins of Tui Regio.
If these structures are deltas, it could imply that the channels supplied the paleo-sea. One ice-rich (high $\delta$) area (130$\pm$1$^o$W,16$\pm$1.5$^o$S) does not seem to end a channel, but it may be because it was too old or too narrow to be detected on the RADAR map. 
If precipitations are sufficient, rainstorms also could make small or subsidiary channels not resolved with the RADAR resolution overflows and deposit washed off sediments at the margins of the paleo-sea.

Another ice-rich area (119$\pm$2$^o$W, 24$\pm$1.$^o$S) 
seems to be connected to a remote channel through a broad and diffuse river bed, not different but much longer that the structures that connects the other channels to the ice-rich zones. This is a possible old channel, highlighted in Fig.~\ref{ratio2} (in purple dotted line). 

Our results also show a gradual increase of $\delta$ from the center towards the borders of Tui Regio. This result is consistent with the pattern predicted by \citet{cordier_etal_2016b}, in their Fig.~10 showing a decrease of the evaporitic layer's thickness with the distance from the center of the sea. However VIMS only probes the surface firsts tens of microns, so if there is a thickness variation, it is on a micrometric scale. A more probable interpretation would be a change of composition in the evaporitic layer due to the different saturation values of the components in the liquid \citep{cordier_etal_2013b,cordier_etal_2016b}.

North to the deltas, depending on the observation, color on Xanadu mountains are contrasted. In C15, Xanadu mountains appear with a high ice-index (forming the ice-rich strip discussed previously). The 
plain at the foot of the mountains has a low ice signal and it is then probably covered by sediment. In C18, Xanadu mountains and the plains below appear to be quite uniform, regardless the topography. Differences between the two observations, as explained above, may be due to meteorological events or to the different incident and emergent angles if the surface reflectivity is not isotropic as assumed in our model.\\
The explanation of our ice index maps (ratios and $\delta$) with dry (occasionnally active) river beds ending in delta formations at the edge of a paleolake appears the most consistent with observations. These rivers seem to flow from Xanadu mountain slopes. The site topology could have been formed by impact craters but they do not seems to be the direct reason for ice rich terrains to be apparent.  

%\subsection{Lake}
%The dry lac at 25$^{\circ}$S118$^{\circ}$W may be a complex Sharp-Edged Depression (SED) \citep{hayes_etal_2017}, as we can distinguish a ridge at the borders on the SAR map in Fig.~\ref{radarVIMS}. As study of the detailed topography is necessary to confirm this assumption. It could mean that those SEDs don't only form at the poles. We find an extremely low water-ice signature in this lake, which is coherent with an evaporites formations, and a sedimentation of Titan's aerosols.

\section{Conclusion}

In this work we used a model of photometry to account for atmosphere scattering and absorption. The purpose is to retrieve the surface albedo and to remove as much as possible the effects due to the atmosphere and the observation geometries. The model setup is quite standard and derived from previous models  \citep{rannou_etal_2010}. Among other specific characteristics, we use a model of scattering by fractal aggregates \citep{Rannou1997} to calculate the optical properties of the haze. Our simulations show the need for aerosols with a fractal dimension between 2.3 and 2.4, i.e., more compact than initially assumed (2.0). This is obviously meaningful in term of aerosol growth and in term of mechanical behavior in the atmosphere. This updated fractal dimension reduces and flattens the surface albedo spectra. It also changes its shape at 1.3 $\mu$m. %This change arised with recent calibrations of VIMS data, the will to improve the fit around 1 and 1.8 $\mu$m, and the modifications brought by the updated haze and mist vertical profiles. 

The haze opacity varies over large distances, and not always along the latitude. The mist opacity factor changes over shorter distances than the haze opacity factor, as expected since it is more sensitive to variations of the lower atmosphere. Our study confirms that these two factors are not correlated. Seasonal cycle may also increase the haze parameter, but further studies are required to confirm the influence of a seasonal cycle on the haze over several years.

The photometry model enabled us to retrieve the surface albedo in seven methane windows, and for the pixels of two VIMS cubes probing the region between Xanadu and Tui regio. This region is of some interest as it is assumed to be an ancient sea. We established several criteria to detect water ice signature (four ratios and a global index). The radiative transfer analysis highlights areas of high water-ice signal on Xanadu mountains and also at the far end of dry channels. Areas shapes of high water-ice signal at the stream mouths are triangular as expected for alluvial fans, or deposits of deltas. In the first case an abrupt  slope change is needed, and in the second case the channels supply the paleo-sea of Tui Regio. Impact craters with ice-rich rims is a less likely explication for these features. The ice signal decreases inside the channels, perhaps due to their sizes, smaller than the spatial resolution of a VIMS pixel. Anyhow, our results support \citet{barnes_etal_2007}, \citet{langhans_etal_2012} and \citet{brossier2018} hypothesis of sediment transport from Xanadu mountains to the channels downstream, with a decreasing grain size. 

In addition, our analysis shows a gradient of low $\delta$ perfectly superposed to Tui Regio. This gradient could correspond to the decreasing thickness from the center to the border of the evaporite layer of Tui Regio, as suggested by \citet{cordier_etal_2016b}, or a variation of composition in the evaporite layer.
Here the overall shape of the albedo between a low and high $\delta$ does not really change, except at 2.8 $\mu$m, i.e., in a domain where the model is incomplete.
Detailed analysis of the retrieved albedo could bring further insight into the spectral composition of these terrains provided the uncertainties are small enough. We used an original method to provide estimate of the VIMS instrument intrinsic noise. This latter tends to increase with the wavelength, with $\sim$ 1\% at short wavelengths to 10-40\% above 3.5$\mu$m. The error also varies from one cube to another.

Comparing our results to previous global mapping and RGB composites (Fig.~\ref{Barnes}), we see that the link between water ice and 'dark blue', and 'bright blue' areas is not as straightforward as it was previously thought \citep{Barnes2007b}. The ice signature can be modulated by grain size, dust cover that darkens the albedo, or even the incident angles. Indeed, a light blue area in Fig.~\ref{Barnes} is actually associated with a low water ice signal in Fig.~\ref{ratio2}-a whereas some black units corresponds to high values of $\delta$. Only the RT analysis shows qualitatively the difference in intensity of water-ice signature between those blue units.

A surface albedo map leads to a better understanding of the structures. It allows removal of the atmospheric inputs and to correct effects of different observation geometries - except for the surface photometric function since we model surface reflectivity as a lambertian surface. These effects can deeply alter the perception of the true surface features by inducing strong biases. The different albedo ratios and the parameter $\delta$ that we consider in this work also have a much better contrasts than a direct analysis of radiance at the top of the atmosphere. They give a direct appreciation of the strength of the water-ice signature. Nonetheless, we are conscious of the limits of $\delta$. Since it is calculated using normalized ratios, meaning that for a VIMS cube with absolutely no water-ice, $\delta$ could still reach 0.5. That is why we must use it in addition to other methods of surface analysis. The model can be improved to obtain identical absolute ratios for VIMS cubes in a similar area, at the same time period and under different geometries. This is possible, for example, with better aerosol phase functions and better information about the vertical profile of haze and mist.
More study is required on the mist and haze to retrieve the absolute value of the contrasts observed as water-ice signature. If our model is not perfect, it shows a possible way to obtain more information about the surface in the future.

The terrains studied in this paper are of great interest in terms of exobiology. Similar terrains on Earth, with sediment accumulation areas (deltas, paleo-lakes) are places where bio-signatures are generally well preserved \citep{meyers2002}. It would be particularly relevant for \textit{Dragonfly} to visit areas similar to the region studied in our work \citep{lorenz_etal_2018b}. The mission will target zones with ice-rich material, which will help alleviate the ambiguity of VIMS analysis by finding the composition of the famous 'dark material' often mixed with the ice \citep{barnes2020}.

There are probably plenty of other areas like those identified but not resolved by VIMS: the images from the descent of \textit{Huygens} actually suggest the existence of river networks that are undetectable by VIMS or the RADAR because of their small dimensions \citep{owen2005}.

Lastly, the approach proposed here can be applied to other regions in order to study qualitatively the water-ice signature on Titan, like Shangri-La, future landing site of \textit{Dragonfly} and where \textit{Huygens} landed. The averred flooded area in \citet{turtle_eal_2011a} paper could also be of interest. The model needs no specific adaptation for equatorial or tropical regions, but probably moderate adaptation for mid-latitude regions (from $\pm$ 30$^o$ to 50$^o$) because the haze layer still has a vertical profile similar to the equatorial one. On the other hand, to be able to study polar regions, we need to adapt the vertical profile of aerosols and mist, and likely their spectral behavior.

\trait

\section*{Data availability and final version}
The RT analysis results on C14, C15 and C18 are available here: https://doi.org/10.5281/zenodo.4675563. The final and peer-reviewed version of this paper is available here: https://doi.org/10.1016/j.icarus.2021.114464.

\trait

\section*{Acknowledgements}
The present research was supported by the Programme National de Plan\'{e}tologie (PNP) of CNRS-INSU co-funded by CNES, and also
supported by the French HPC Center ROMEO, the R\'{e}gion Grand-Est and the University of Reims Champagne-Ardenne. Many thanks to J\'{e}r\'{e}mie Burgalat for the IT help provided.

%%%%%%%%%%%%%%%%%%%%%%%%%%%%%%%%%%%%%%%%%%%%%%%%%%%%%%%%%%%%%%%%%%%%%%%%%%%%%%%%%%%%%%%%%%%%%%%%%%%%%%%%%%%%%%%%%%%%%%%%%%%%%%%%%%%%
%% NOTES :

%\clearpage
%\section{NOTES -- IMPORTANT}
%\vspace{1cm}

%\noindent{\color{CNRSBlue}{\bf  : }
%\begin{itemize}
  %\item[\textbullet] .\\
  %\item[\textbullet] 
%\end{itemize}
%}

%\noindent{\color{CNRSBlue}{\bf Des elements envoyes par Maelie le 08/10/2019 :}\\

%{\color{CNRSBlue}
%Du to the methane absorptions in the thick atmosphere of Titan, the surface can only be observed in $7$ atmospheric windows centred at $0.93$, $1.08$, $1.27$, $1.59$, $2.01$, $2.7-2.8$ and $5$~$\mu$m \citep{sotin_etal_2005,barnes_etal_2005}. The limits of the windows are defined with the radiative transfer model, observing the differences in modelized spectra when we change the surface albedo.}\\

\clearpage
\bibliographystyle{apj}
\bibliography{./bibliographie_planeto_2019,./biblio_Maelie}

\def\sciam{Sci.
  Am.}\def\nature{Nature}\def\nat{Nature}\def\science{Science}\def\natastro{Nat.
  Astron.}\def\natgeo{Nat. Geosci.}\def\natcom{Nat.
  Commun.}\def\pnas{PNAS}\def\AnnderPhys{‎Ann. Phys.
  (Berl.)}\def\icarus{Icarus}\def\pss{Planet. Space Sci.}\def\planss{Planet.
  Space Sci.}\def\ssr{Space Sci. Rev.}\def\solsr{Sol. Syst. Res.}\def\jqsrt{J.
  Quant. Spectrosc. Radiat. Transfer}\def\expastro{Exp. Astron.}\def\jcis{‎J.
  Colloid Interface
  Sci.}\def\aap{A\&A}\def\apj{ApJ}\def\apjl{ApJL}\def\apjs{ApJS}\def\aj{AJ}\def\mnras{MNRAS}\def\araa{Annu.
  Rev. Astron. Astrophys.}\def\pasj{Publ. Astron. Soc.
  Jpn.}\def\apss{Astrophys. Space Sci.}\def\pasp{Publ. Astron. Soc.
  Pac.}\def\expastron{Exp. Astron.}\def\asr{Adv. Space
  Res.}\def\astrobiol{Astrobiology}\def\areps{Annu. Rev. Earth Planet.
  Sci.}\def\georl{Geophys. Res. Lett.}\def\grl{Geophys. Res. Lett.}\def\jgr{J.
  Geophys. Res.}\def\gca{Geochim. Cosmochim. Ac.}\def\epsl{Earth Planet. Sci.
  Lett.}\def\plasci{Planet. Sci.}\def\ggg{Geochem. Geophys.
  Geosyst.}\def\rmg{Rev. Mineral. Geochem.}\def\tpm{Transport Porous
  Med.}\def\philtrans{Phil. Trans.}\def\faradis{Farad. Discuss.}\def\jcis{‎J.
  Colloid Interface Sci.}\def\jfm{J. Fluid Mech.}\def\physflu{Phys.
  Fluids}\def\pachem{Pure Appl. Chem.}\def\jpcA{J. Phys. Chem.
  A}\def\chemrev{Chem. Rev.}\def\AppOpt{Appl.
  Opt.}\def\nature{Nature}\def\nat{Nature}\def\science{Science}\def\natastro{Nat.
  Astron.}\def\natgeo{Nat. Geosci.}\def\natcom{Nat. Commun.}\def\scirep{Sci.
  Rep.}\def\science{Sci}\def\jced{J. Chem. Eng. Data}\def\fpe{Fluid Phase
  Equilibria}\def\iecr{Ind. Eng. Chem. Res.}\def\aichej{AIChE J.}\def\pt{Powder
  Technol.}\def\etfs{Exp. Therm. Fluid Sci.}\def\jgr{J. Geophys.
  Res.}\def\gca{Geochim. Cosmochim. Acta}\def\chemgeol{Chem Geol.}\def\jcp{J.
  Chem. Phys.}\def\jcis{‎J. Colloid Interface Sci.}\def\jcsft{J. Chem. Soc.
  Faraday Trans.}\def\jpcB{J. Phys. Chem. B}\def\jsf{J. Supercrit.
  Fluids}\def\enerp{Energy Procedia}\def\aichej{AlChE J.}\def\IECPDD{Ind. Eng.
  Chem. Process Des. Dev.}\def\EF{Energy Fuels}\def\jacs{J. Am. Chem. Soc.}
\begin{thebibliography}{}
\expandafter\ifx\csname natexlab\endcsname\relax\def\natexlab#1{#1}\fi

\bibitem[{Annex {et~al.}(2020)Annex, Pearson, Seignovert, Carcich, Eichhorn,
  Mapel, von Forstner, McAuliffe, del Rio, Berry, Aye, Stefko, de~Val-Borro,
  Kulumani, \& ya~Murakami}]{Annex2020}
Annex, A.~M., Pearson, B., Seignovert, B., {et~al.} 2020, Journal of Open
  Source Software, 5, 2050

\bibitem[{Barnes {et~al.}(2020)Barnes, Turtle, Trainer, Lorenz, Murchie, \&
  MacKenzie}]{barnes2020}
Barnes, J., Turtle, E., Trainer, M., {et~al.} 2020, in American Astronomical
  Society Meeting Abstracts\# 236, Vol. 236, 221--05

\bibitem[{{Barnes} {et~al.}(2005){Barnes}, {Brown}, {Turtle}, {McEwen},
  {Lorenz}, {Janssen}, {Schaller}, {Brown}, {Buratti}, {Sotin}, {Griffith},
  {Clark}, {Perry}, {Fussner}, {Barbara}, {West}, {Elachi}, {Bouchez}, {Roe},
  {Baines}, {Bellucci}, {Bibring}, {Capaccioni}, {Cerroni}, {Combes},
  {Coradini}, {Cruikshank}, {Drossart}, {Formisano}, {Jaumann}, {Langevin},
  {Matson}, {McCord}, {Nicholson}, \& {Sicardy}}]{barnes_etal_2005}
{Barnes}, J.~W., {Brown}, R.~H., {Turtle}, E.~P., {et~al.} 2005, Science, 310,
  92

\bibitem[{Barnes {et~al.}(2007)Barnes, Brown, Soderblom, Buratti, Sotin,
  Rodriguez, Mouèlic], Baines, Clark, \& Nicholson}]{Barnes2007b}
Barnes, J.~W., Brown, R.~H., Soderblom, L., {et~al.} 2007, Icarus, 186, 242

\bibitem[{{Barnes} {et~al.}(2007){Barnes}, {Radebaugh}, {Brown}, {Wall},
  {Soderblom}, {Lunine}, {Burr}, {Sotin}, {Le Mou{\'e}lic}, {Rodriguez},
  {Buratti}, {Clark}, {Baines}, {Jaumann}, {Nicholson}, {Kirk}, {Lopes},
  {Lorenz}, {Mitchell}, \& {Wood}}]{barnes_etal_2007}
{Barnes}, J.~W., {Radebaugh}, J., {Brown}, R.~H., {et~al.} 2007, Journal of
  Geophysical Research (Planets), 112, E11006

\bibitem[{Barnes {et~al.}(2008)Barnes, Brown, Soderblom, Sotin, Le~Mou{\`e}lic,
  Rodriguez, Jaumann, Beyer, Buratti, Pitman, {et~al.}}]{barnes2008}
Barnes, J.~W., Brown, R.~H., Soderblom, L., {et~al.} 2008, Icarus, 195, 400

\bibitem[{{Barnes} {et~al.}(2011){Barnes}, {Bow}, {Schwartz}, {Brown},
  {Soderblom}, {Hayes}, {Vixie}, {Le Mou{\'e}lic}, {Rodriguez}, {Sotin},
  {Jaumann}, {Stephan}, {Soderblom}, {Clark}, {Buratti}, {Baines}, \&
  {Nicholson}}]{barnes_etal_2011}
{Barnes}, J.~W., {Bow}, J., {Schwartz}, J., {et~al.} 2011, \icarus, 216, 136

\bibitem[{Barnes {et~al.}(2018)Barnes, MacKenzie, Young, Trouille, Rodriguez,
  Cornet, Jackson, {\'{A}}d{\'{a}}mkovics, Sotin, \& Soderblom}]{Barnes2018}
Barnes, J.~W., MacKenzie, S.~M., Young, E.~F., {et~al.} 2018, The Astronomical
  Journal, 155, 264

\bibitem[{Birch {et~al.}(2016)Birch, Hayes, Howard, Moore, \&
  Radebaugh}]{Birch2016}
Birch, S., Hayes, A., Howard, A., Moore, J., \& Radebaugh, J. 2016, Icarus,
  270, 238 , titan's Surface and Atmosphere

\bibitem[{Bonnefoy {et~al.}(2016)Bonnefoy, Hayes, Hayne, Malaska, {Le Gall},
  Solomonidou, \& Lucas}]{Bonnefoy2016}
Bonnefoy, L.~E., Hayes, A.~G., Hayne, P.~O., {et~al.} 2016, Icarus, 270, 222 ,
  titan's Surface and Atmosphere

\bibitem[{Botet {et~al.}(1995)Botet, Rannou, \& Cabane}]{Botet_1995}
Botet, R., Rannou, P., \& Cabane, M. 1995, Journal of Physics A: Mathematical
  and General, 28, 297

\bibitem[{Brossier {et~al.}(2018)Brossier, Rodriguez, Cornet, Lucas, Radebaugh,
  Maltagliati, Le~Mou{\'e}lic, Solomonidou, Coustenis, Hirtzig,
  {et~al.}}]{brossier2018}
Brossier, J., Rodriguez, S., Cornet, T., {et~al.} 2018, Journal of Geophysical
  Research: Planets, 123, 1089

\bibitem[{Brown {et~al.}(2011)Brown, Barnes, \& Melosh}]{Brown2011}
Brown, R.~H., Barnes, J.~W., \& Melosh, H.~J. 2011, Icarus, 214, 556

\bibitem[{{Brown} {et~al.}(2004){Brown}, {Baines}, {Bellucci}, {Bibring},
  {Buratti}, {Capaccioni}, {Cerroni}, {Clark}, {Coradini}, {Cruikshank},
  {Drossart}, {Formisano}, {Jaumann}, {Langevin}, {Matson}, {McCord},
  {Mennella}, {Miller}, {Nelson}, {Nicholson}, {Sicardy}, \&
  {Sotin}}]{brown_etal_2004}
{Brown}, R.~H., {Baines}, K.~H., {Bellucci}, G., {et~al.} 2004, ssr, 115, 111

\bibitem[{Charnay {et~al.}(2015)Charnay, Barth, Rafkin, Narteau, Lebonnois,
  Rodriguez, Du~Pont, \& Lucas}]{charnay2015}
Charnay, B., Barth, E., Rafkin, S., {et~al.} 2015, Nature Geoscience, 8, 362

\bibitem[{{Clark} {et~al.}(2018){Clark}, {Brown}, {Lytle}, \&
  {Hedman}}]{clark_etal_2018}
{Clark}, R.~N., {Brown}, R.~H., {Lytle}, D.~M., \& {Hedman}, M. 2018, NASA
  Planetary Data System, The Planetary Atmospheres Node

\bibitem[{{Cordier} {et~al.}(2013){Cordier}, {Barnes}, \&
  {Ferreira}}]{cordier_etal_2013b}
{Cordier}, D., {Barnes}, J.~W., \& {Ferreira}, A.~G. 2013, \icarus, 226, 1431

\bibitem[{{Cordier} {et~al.}(2016){Cordier}, {Cornet}, {Barnes}, {MacKenzie},
  {Le Bahers}, {Nna Mvondo}, \& {Ferreira}}]{cordier_etal_2016b}
{Cordier}, D., {Cornet}, T., {Barnes}, J.~W., {et~al.} 2016, \icarus, 270,
  41–56

\bibitem[{{Corlies} {et~al.}(2017){Corlies}, {Hayes}, {Birch}, {Lorenz},
  {Stiles}, {Kirk}, {Poggiali}, {Zebker}, \& {Iess}}]{corlies_etal_2017}
{Corlies}, P., {Hayes}, A.~G., {Birch}, S.~P.~D., {et~al.} 2017, \georl, 44, 11

\bibitem[{Cours {et~al.}(2020)Cours, Cordier, Seignovert, Maltagliati, \&
  Biennier}]{Cours2020}
Cours, T., Cordier, D., Seignovert, B., Maltagliati, L., \& Biennier, L. 2020,
  Icarus, 339, 113571

\bibitem[{de~Bergh {et~al.}(2012)de~Bergh, Courtin, Bézard, Coustenis,
  Lellouch, Hirtzig, Rannou, Drossart, Campargue, Kassi, Wang, Boudon, Nikitin,
  \& Tyuterev}]{DEBERGH2012}
de~Bergh, C., Courtin, R., Bézard, B., {et~al.} 2012, Planetary and Space
  Science, 61, 85 , surfaces, atmospheres and magnetospheres of the outer
  planets and their satellites and ring systems: Part VII

\bibitem[{Doose {et~al.}(2016)Doose, Karkoschka, Tomasko, \&
  Anderson}]{Doose2016}
Doose, L.~R., Karkoschka, E., Tomasko, M.~G., \& Anderson, C.~M. 2016, Icarus,
  270, 355 , titan's Surface and Atmosphere

\bibitem[{Emde {et~al.}(2016)Emde, Buras-Schnell, Kylling, Mayer, Gasteiger,
  Hamann, Kylling, Richter, Pause, Dowling, \& Bugliaro}]{libRadtran2016}
Emde, C., Buras-Schnell, R., Kylling, A., {et~al.} 2016, Geoscientific Model
  Development, 9, 1647

\bibitem[{{Evans}(2007)}]{evans_2007}
{Evans}, K.~F. 2007, J. Atmos. Sci., 64, 3854

\bibitem[{Faulk {et~al.}(2017)Faulk, Mitchell, Moon, \& Lora}]{Faulk2017}
Faulk, S.~P., Mitchell, J.~L., Moon, S., \& Lora, J.~M. 2017, Nature
  Geoscience, 10, 827

\bibitem[{{Fulchignoni} {et~al.}(2005){Fulchignoni}, {Ferri}, {Angrilli},
  {Ball}, {Bar-Nun}, {Barucci}, {Bettanini}, {Bianchini}, {Borucki},
  {Colombatti}, {Coradini}, {Coustenis}, {Debei}, {Falkner}, {Fanti},
  {Flamini}, {Gaborit}, {Grard}, {Hamelin}, {Harri}, {Hathi}, {Jernej},
  {Leese}, {Lehto}, {Lion Stoppato}, {L{\'o}pez-Moreno}, {M{\"a}kinen},
  {McDonnell}, {McKay}, {Molina-Cuberos}, {Neubauer}, {Pirronello}, {Rodrigo},
  {Saggin}, {Schwingenschuh}, {Seiff}, {Sim{\~o}es}, {Svedhem}, {Tokano},
  {Towner}, {Trautner}, {Withers}, \& {Zarnecki}}]{fulchignoni_etal_2005}
{Fulchignoni}, M., {Ferri}, F., {Angrilli}, F., {et~al.} 2005, \nat, 438, 785

\bibitem[{{Goody} {et~al.}(1989){Goody}, {West}, {Chen}, \&
  {Crisp}}]{goody_etal_1989}
{Goody}, R., {West}, R., {Chen}, L., \& {Crisp}, D. 1989, \jqsrt, 42, 539

\bibitem[{{Griffith} {et~al.}(2012){Griffith}, {Lora}, {Turner}, {Penteado},
  {Brown}, {Tomasko}, {Doose}, \& {See}}]{griffith_etal_2012}
{Griffith}, C.~A., {Lora}, J.~M., {Turner}, J., {et~al.} 2012, \nat, 486, 237

\bibitem[{{Griffith} {et~al.}(2019){Griffith}, {Penteado}, {Turner}, {Neish},
  {Mitri}, {Montiel}, {Schoenfeld}, \& {Lopes}}]{griffith_etal_2019}
{Griffith}, C.~A., {Penteado}, P.~F., {Turner}, J.~D., {et~al.} 2019,
  \natastro, 343

\bibitem[{{Griffith} {et~al.}(2005){Griffith}, {Penteado}, {Baines},
  {Drossart}, {Barnes}, {Bellucci}, {Bibring}, {Brown}, {Buratti},
  {Capaccioni}, {Cerroni}, {Clark}, {Combes}, {Coradini}, {Cruikshank},
  {Formisano}, {Jaumann}, {Langevin}, {Matson}, {McCord}, {Mennella}, {Nelson},
  {Nicholson}, {Sicardy}, {Sotin}, {Soderblom}, \&
  {Kursinski}}]{griffith_etal_2005}
{Griffith}, C.~A., {Penteado}, P., {Baines}, K., {et~al.} 2005, \science, 310,
  474

\bibitem[{Gyr \& Hoyer(2006)}]{Sediment_Transport}
Gyr, A., \& Hoyer, K. 2006, Sediment Transport, 1st edn. (Springer Netherlands)

\bibitem[{Hansen \& Travis(1974)}]{Hansen1974}
Hansen, J.~E., \& Travis, L.~D. 1974, Space Science Reviews, 16, 527

\bibitem[{{Hofgartner} {et~al.}(2020){Hofgartner}, {Hayes}, {Campbell},
  {Lunine}, {Black}, {MacKenzie}, {Birch}, {Elachi}, {Kirk}, {Le Gall},
  {Lorenz}, \& {Wall}}]{hofgartner_etal_2020}
{Hofgartner}, J.~D., {Hayes}, A.~G., {Campbell}, D.~B., {et~al.} 2020, \natcom,
  11, 2829

\bibitem[{Janssen {et~al.}(2016)Janssen, Gall, Lopes, Lorenz, Malaska, Hayes,
  Neish, Solomonidou, Mitchell, Radebaugh, Keihm, Choukroun, Leyrat, Encrenaz,
  \& Mastrogiuseppe}]{Janssen2016}
Janssen, M., Gall, A.~L., Lopes, R., {et~al.} 2016, Icarus, 270, 443 , titan's
  Surface and Atmosphere

\bibitem[{{Karkoschka}(2016)}]{karkoschka_2016}
{Karkoschka}, E. 2016, \icarus, 270, 339

\bibitem[{Karkoschka {et~al.}(2012)Karkoschka, Schr{\"o}der, Tomasko, \&
  Keller}]{karkoschka2012}
Karkoschka, E., Schr{\"o}der, S.~E., Tomasko, M.~G., \& Keller, H.~U. 2012,
  Planetary and Space Science, 60, 342

\bibitem[{Kazeminejad {et~al.}(2011)Kazeminejad, Atkinson, \&
  Lebreton}]{Kazeminejad2011}
Kazeminejad, B., Atkinson, D.~H., \& Lebreton, J.-P. 2011, Advances in Space
  Research, 47, 1622

\bibitem[{Lafferty {et~al.}(1996)Lafferty, Solodov, Weber, Olson, \&
  Hartmann}]{Lafferty1996}
Lafferty, W.~J., Solodov, A.~M., Weber, A., Olson, W.~B., \& Hartmann, J.-M.
  1996, Appl. Opt., 35, 5911

\bibitem[{{Langhans} {et~al.}(2012){Langhans}, {Jaumann}, {Stephan}, {Brown},
  {Buratti}, {Clark}, {Baines}, {Nicholson}, {Lorenz}, {Soderblom},
  {Soderblom}, {Sotin}, {Barnes}, \& {Nelson}}]{langhans_etal_2012}
{Langhans}, M.~H., {Jaumann}, R., {Stephan}, K., {et~al.} 2012, \planss, 60, 34

\bibitem[{{Le Gall} {et~al.}(2010){Le Gall}, {Janssen}, {Paillou}, {Lorenz},
  {Wall}, \& {the Cassini Radar Team}}]{legall_etal_2010}
{Le Gall}, A., {Janssen}, M.~A., {Paillou}, P., {et~al.} 2010, \icarus, 207,
  948

\bibitem[{{Le Mou{\'e}lic} {et~al.}(2019){Le Mou{\'e}lic}, {Cornet},
  {Rodriguez}, {Sotin}, {Seignovert}, {Barnes}, {Brown}, {Baines}, {Buratti},
  {Clark}, {Nicholson}, {Lasue}, {Pasek}, \& {Soderblom}}]{lemouelic_etal_2019}
{Le Mou{\'e}lic}, S., {Cornet}, T., {Rodriguez}, S., {et~al.} 2019, \icarus,
  319, 121

\bibitem[{Lellouch {et~al.}(2003)Lellouch, Coustenis, Sebag, Cuby,
  López-Valverde, Schmitt, Fouchet, \& Crovisier}]{Lellouch2003}
Lellouch, E., Coustenis, A., Sebag, B., {et~al.} 2003, Icarus, 162, 125

\bibitem[{{Levenberg}(1944)}]{levenberg_1944}
{Levenberg}, K. 1944, Quarterly of Applied Mathematic, 2, 164

\bibitem[{Lopes {et~al.}(2019)Lopes, Malaska, Schoenfeld, Solomonidou, Birch,
  Florence, Hayes, Williams, Radebaugh, Verlander, Turtle, Le~Gall, \&
  Wall}]{Lopes2019}
Lopes, R., Malaska, M., Schoenfeld, A., {et~al.} 2019, Nature Astronomy, 1

\bibitem[{{Lorenz} {et~al.}(2018){Lorenz}, {Turtle}, {Barnes}, ~, {Adams},
  {Hibbard}, {Sheldon}, {Zacny}, {Peplowski}, {Lawrence}, {Ravine}, {McGee},
  {Sotzen}, {MacKenzie}, {Langelaan}, {Schmitz}, {Wolfarth}, \&
  {Bedini}}]{lorenz_etal_2018b}
{Lorenz}, R.~D., {Turtle}, E.~P., {Barnes}, J.~W., {et~al.} 2018, {Dragonfly: A
  Rotorcraft Lander Concept for Scientific Exploration at Titan}, Tech. Rep.~3,
  Johns Hopkins APL

\bibitem[{MacKenzie \& Barnes(2016)}]{MacKenzie2016}
MacKenzie, S., \& Barnes, J. 2016, The Astrophysical Journal, 821, 0

\bibitem[{{MacKenzie} {et~al.}(2014){MacKenzie}, {Barnes}, {Sotin},
  {Soderblom}, {Le Mou{\'e}lic}, {Rodriguez}, {Baines}, {Buratti}, {Clark},
  {Nicholson}, \& {McCord}}]{macKenzie_etal_2014}
{MacKenzie}, S.~M., {Barnes}, J.~W., {Sotin}, C., {et~al.} 2014, \icarus, 243,
  191

\bibitem[{Malaska {et~al.}(2016)Malaska, Lopes, Williams, Neish, Solomonidou,
  Soderblom, Schoenfeld, Birch, Hayes, Gall, Janssen, Farr, Lorenz, Radebaugh,
  \& Turtle}]{malaska2016}
Malaska, M.~J., Lopes, R.~M., Williams, D.~A., {et~al.} 2016, Icarus, 270, 130
  , titan's Surface and Atmosphere

\bibitem[{Maltagliati {et~al.}(2015)Maltagliati, Bézard, Vinatier, Hedman,
  Lellouch, Nicholson, Sotin, [de Kok], \& Sicardy}]{Maltagliati2015}
Maltagliati, L., Bézard, B., Vinatier, S., {et~al.} 2015, Icarus, 248, 1

\bibitem[{{Marquardt}(1963)}]{marquardt_1963}
{Marquardt}, D. 1963, SIAM Journal on Applied Mathematics, 11, 431

\bibitem[{Mayer \& Kylling(2005)}]{mayer2005}
Mayer, B., \& Kylling, A. 2005, {Atmospheric Chemistry and Physics
  Discussions}, 5, 1319

\bibitem[{{McCord} {et~al.}(2006){McCord}, {Hansen}, {Buratti}, {Clark},
  {Cruikshank}, {D'Aversa}, {Griffith}, {Baines}, {Brown}, {Dalle Ore},
  {Filacchione}, {Formisano}, {Hibbitts}, {Jaumann}, {Lunine}, {Nelson},
  {Sotin}, \& {the Cassini VIMS Team}}]{mccord_etal_2006}
{McCord}, T.~B., {Hansen}, G.~B., {Buratti}, B.~J., {et~al.} 2006, \planss, 54,
  1524

\bibitem[{{McKay} {et~al.}(1989){McKay}, {Pollack}, \&
  {Courtin}}]{mckay_etal_1989}
{McKay}, C.~P., {Pollack}, J.~B., \& {Courtin}, R. 1989, \icarus, 80, 23

\bibitem[{McKellar(1989)}]{mckellar1989}
McKellar, A. 1989, Icarus, 80, 361

\bibitem[{Meyers \& Teranes(2002)}]{meyers2002}
Meyers, P.~A., \& Teranes, J.~L. 2002, in Tracking environmental change using
  lake sediments (Springer), 239--269

\bibitem[{Moore \& Howard(2010)}]{Moore2010}
Moore, J.~M., \& Howard, A.~D. 2010, Geophysical Research Letters, 37,
  https://agupubs.onlinelibrary.wiley.com/doi/pdf/10.1029/2010GL045234

\bibitem[{{Mor\'{e}} {et~al.}(1980){Mor\'{e}}, {Garbow}, \&
  {Hillstrom}}]{more_minpack}
{Mor\'{e}}, J.~J., {Garbow}, B.~S., \& {Hillstrom}, K.~E. 1980, {User Guide for
  MINPACK-1}, Tech. Rep. ANL-80-74, pub-ANL, pub-ANL:adr

\bibitem[{Neish {et~al.}(2013)Neish, Kirk, Lorenz, Bray, Schenk, Stiles,
  Turtle, Mitchell, \& Hayes}]{Neish2013}
Neish, C.~D., Kirk, R.~L., Lorenz, R.~D., {et~al.} 2013, Icarus, 223, 82

\bibitem[{Nicholson {et~al.}(2020)Nicholson, Ansty, Hedman, Creel, Ahlers,
  Harbison, Brown, Clark, Baines, Buratti, Sotin, \& Badman}]{Nicholson2020}
Nicholson, P.~D., Ansty, T., Hedman, M.~M., {et~al.} 2020, Icarus, 344, 113356,
  cassini Mission Science Results

\bibitem[{{Niemann} {et~al.}(2005){Niemann}, {Atreya}, {Bauer}, {Carignan},
  {Demick}, {Frost}, {Gautier}, {Haberman}, {Harpold}, {Hunten}, {Israel},
  {Lunine}, {Kasprzak}, {Owen}, {Paulkovich}, {Raulin}, {Raaen}, \&
  {Way}}]{niemann_etal_2005}
{Niemann}, H.~B., {Atreya}, S.~K., {Bauer}, S.~J., {et~al.} 2005, \nature, 438,
  779

\bibitem[{{Niemann} {et~al.}(2010){Niemann}, {Atreya}, {Demick}, {Gautier},
  {Haberman}, {Harpold}, {Kasprzak}, {Lunine}, {Owen}, \&
  {Raulin}}]{niemann_etal_2010}
{Niemann}, H.~B., {Atreya}, S.~K., {Demick}, J.~E., {et~al.} 2010, \jgr, 115,
  E12006

\bibitem[{Nixon {et~al.}(2012)Nixon, Temelso, Vinatier, Teanby, B{\'{e}}zard,
  Achterberg, Mandt, Sherrill, Irwin, Jennings, Romani, Coustenis, \&
  Flasar}]{Nixon_2012}
Nixon, C.~A., Temelso, B., Vinatier, S., {et~al.} 2012, The Astrophysical
  Journal, 749, 159

\bibitem[{Owen(2005)}]{owen2005}
Owen, T. 2005, Nature, 438, 756

\bibitem[{Radebaugh {et~al.}(2018)Radebaugh, Ventra, Lorenz, Farr, Kirk, Hayes,
  Malaska, Birch, Liu, Lunine, {et~al.}}]{radebaugh2018}
Radebaugh, J., Ventra, D., Lorenz, R.~D., {et~al.} 2018, Geological Society,
  London, Special Publications, 440, 281

\bibitem[{Rannou {et~al.}(1997)Rannou, Cabane, Botet, \&
  Chassefière}]{Rannou1997}
Rannou, P., Cabane, M., Botet, R., \& Chassefière, E. 1997, Journal of
  Geophysical Research: Planets, 102, 10997

\bibitem[{{Rannou} {et~al.}(2010){Rannou}, {Cours}, {Le Mou{\'e}lic},
  {Rodriguez}, {Sotin}, {Drossart}, \& {Brown}}]{rannou_etal_2010}
{Rannou}, P., {Cours}, T., {Le Mou{\'e}lic}, S., {et~al.} 2010, \icarus, 208,
  850

\bibitem[{{Rannou} {et~al.}(2004){Rannou}, {Hourdin}, {McKay}, \&
  {Luz}}]{rannou_etal_2004}
{Rannou}, P., {Hourdin}, F., {McKay}, C.~P., \& {Luz}, D. 2004, \icarus, 170,
  443

\bibitem[{{Rannou} {et~al.}(2016){Rannou}, {Toledo}, {Lavvas}, {D'Aversa},
  {Moriconi}, {Adriani}, {Le Mou{\'e}lic}, {Sotin}, \&
  {Brown}}]{rannou_etal_2016}
{Rannou}, P., {Toledo}, D., {Lavvas}, P., {et~al.} 2016, \icarus, 270, 291

\bibitem[{{Rey} {et~al.}(2018){Rey}, {Nikitin}, {B\'{e}zard}, {Rannou},
  {Coustenis}, \& {Tyuterev}}]{rey_etal_2018}
{Rey}, M., {Nikitin}, A.~V., {B\'{e}zard}, B., {et~al.} 2018, \icarus, 303, 114

\bibitem[{{Rey} {et~al.}(2016){Rey}, {Nikitin}, \& {Tyuterev}}]{rey_etal_2016}
{Rey}, M., {Nikitin}, A.~V., \& {Tyuterev}, V. 2016, J. Mol. Spectrosc., 327,
  138

\bibitem[{{Rey} {et~al.}(2017){Rey}, {Nikitin}, \& {Tyuterev}}]{rey_etal_2017}
{Rey}, M., {Nikitin}, A.~V., \& {Tyuterev}, V.~G. 2017, \apj, 847, aa8909

\bibitem[{Robidel {et~al.}(2020)Robidel, Mouélic], Tobie, Massé, Seignovert,
  Sotin, \& Rodriguez}]{Robidel2020}
Robidel, R., Mouélic], S.~L., Tobie, G., {et~al.} 2020, Icarus, 349, 113848

\bibitem[{Rodriguez {et~al.}(2006)Rodriguez, Le~Mou{\'e}lic, Sotin, Cl{\'e}net,
  Clark, Buratti, Brown, McCord, Nicholson, Baines, {et~al.}}]{rodriguez2006}
Rodriguez, S., Le~Mou{\'e}lic, S., Sotin, C., {et~al.} 2006, Planetary and
  Space Science, 54, 1510

\bibitem[{Rodriguez {et~al.}(2011)Rodriguez, {Le Mouélic}, Rannou, Sotin,
  Brown, Barnes, Griffith, Burgalat, Baines, Buratti, Clark, \&
  Nicholson}]{Rodriguez2011}
Rodriguez, S., {Le Mouélic}, S., Rannou, P., {et~al.} 2011, Icarus, 216, 89

\bibitem[{Rodriguez {et~al.}(2018)Rodriguez, {Le Mou{\'{e}}lic}, Barnes, Kok,
  Rafkin, Lorenz, Charnay, Radebaugh, Narteau, Cornet, Bourgeois, Lucas,
  Rannou, Griffith, Coustenis, App{\'{e}}r{\'{e}}, Hirtzig, Sotin, Soderblom,
  Brown, Bow, Vixie, Maltagliati, {Courrech du Pont}, Jaumann, Stephan, Baines,
  Buratti, Clark, \& Nicholson}]{Rodriguez2018}
Rodriguez, S., {Le Mou{\'{e}}lic}, S., Barnes, J.~W., {et~al.} 2018, Nature
  Geoscience, 11, 727

\bibitem[{{Rothman} {et~al.}(2013){Rothman}, {Gordon}, {Babikov}, {Barbe},
  {Chris Benner}, {Bernath}, {Birk}, {Bizzocchi}, {Boudon}, {Brown},
  {Campargue}, {Chance}, {Cohen}, {Coudert}, {Devi}, {Drouin}, {Fayt}, {Flaud},
  {Gamache}, {Harrison}, {Hartmann}, {Hill}, {Hodges}, {Jacquemart}, {Jolly},
  {Lamouroux}, {Le Roy}, {Li}, {Long}, {Lyulin}, {Mackie}, {Massie},
  {Mikhailenko}, {M{\"u}ller}, {Naumenko}, {Nikitin}, {Orphal}, {Perevalov},
  {Perrin}, {Polovtseva}, {Richard}, {Smith}, {Starikova}, {Sung}, {Tashkun},
  {Tennyson}, {Toon}, {Tyuterev}, \& {Wagner}}]{rothman_etal_2013}
{Rothman}, L.~S., {Gordon}, I.~E., {Babikov}, Y., {et~al.} 2013, \jqsrt, 130, 4

\bibitem[{{Schr{\"o}der} \& {Keller}(2008)}]{schroder_keller_2008}
{Schr{\"o}der}, S.~E., \& {Keller}, H.~U. 2008, \pss, 56, 753

\bibitem[{Seignovert {et~al.}(2019)Seignovert, Le~Mouélic, Brown, Karkoschka,
  Pasek, Sotin, \& Turtle}]{TitanISSVIMS}
Seignovert, B., Le~Mouélic, S., Brown, R.~H., {et~al.} 2019, Titan’s global
  map combining VIMS and ISS mosaics, doi:10.22002/D1.1173

\bibitem[{Seignovert \& Rannou(2019)}]{Seignovert2019}
Seignovert, B., \& Rannou, P. 2019, PovRay model of fractal aerosols,
  doi:10.5281/zenodo.2559546

\bibitem[{{Seignovert} {et~al.}(2020){Seignovert}, {Rannou}, {West}, \&
  {Vinatier}}]{Seignovert2020}
{Seignovert}, B., {Rannou}, P., {West}, R.~A., \& {Vinatier}, S. 2020, Icarus,
  In revision

\bibitem[{Soderblom {et~al.}(2007)Soderblom, Kirk, Lunine, Anderson, Baines,
  Barnes, Barrett, Brown, Buratti, Clark, Cruikshank, Elachi, Janssen, Jaumann,
  Karkoschka, Mouélic, Lopes, Lorenz, McCord, Nicholson, Radebaugh, Rizk,
  Sotin, Stofan, Sucharski, Tomasko, \& Wall}]{Soderblom2007}
Soderblom, L.~A., Kirk, R.~L., Lunine, J.~I., {et~al.} 2007, Planetary and
  Space Science, 55, 2025 , titan as seen from Huygens

\bibitem[{{Sotin} {et~al.}(2005){Sotin}, {Jaumann}, {Buratti}, {Brown},
  {Clark}, {Soderblom}, {Baines}, {Bellucci}, {Bibring}, {Capaccioni},
  {Cerroni}, {Combes}, {Coradini}, {Cruikshank}, {Drossart}, {Formisano},
  {Langevin}, {Matson}, {McCord}, {Nelson}, {Nicholson}, {Sicardy}, {Le
  Mouelic}, {Rodriguez}, {Stephan}, \& {Scholz}}]{sotin_etal_2005}
{Sotin}, C., {Jaumann}, R., {Buratti}, B.~J., {et~al.} 2005, \nat, 435, 786

\bibitem[{Sromovsky \& Fry(2010)}]{Sromovsky2011}
Sromovsky, L., \& Fry, P. 2010, Icarus, 210, 230

\bibitem[{{Stofan} {et~al.}(2007){Stofan}, {Elachi}, {Lunine}, {Lorenz},
  {Stiles}, {Mitchell}, {Ostro}, {Soderblom}, {Wood}, {Zebker}, {Wall},
  {Janssen}, {Kirk}, {Lopes}, {Paganelli}, {Radebaugh}, {Wye}, {Anderson},
  {Allison}, {Boehmer}, {Callahan}, {Encrenaz}, {Flamini}, {Francescetti},
  {Gim}, {Hamilton}, {Hensley}, {Johnson}, {Kelleher}, {Muhleman}, {Paillou},
  {Picardi}, {Posa}, {Roth}, {Seu}, {Shaffer}, {Vetrella}, \&
  {West}}]{stofan_etal_2007}
{Stofan}, E.~R., {Elachi}, C., {Lunine}, J.~I., {et~al.} 2007, \nature, 445, 61

\bibitem[{{Tomasko} {et~al.}(2008){Tomasko}, {Doose}, {Engel}, {Dafoe}, {West},
  {Lemmon}, {Karkoschka}, \& {See}}]{tomasko_etal_2008b}
{Tomasko}, M.~G., {Doose}, L., {Engel}, S., {et~al.} 2008, \pss, 56, 669

\bibitem[{Turtle {et~al.}(2018)Turtle, Perry, Barbara, Del~Genio, Rodriguez,
  Le~Mou{\'e}lic, Sotin, Lora, Faulk, Corlies, {et~al.}}]{turtle2018}
Turtle, E., Perry, J., Barbara, J., {et~al.} 2018, Geophysical Research
  Letters, 45, 5320

\bibitem[{{Turtle} {et~al.}(2009){Turtle}, {Perry}, {McEwen}, {Del Genio},
  {Barbara}, {West}, {Dawson}, \& {Porco}}]{turtle_etal_2009}
{Turtle}, E.~P., {Perry}, J.~E., {McEwen}, A.~S., {et~al.} 2009, \georl, 36,
  2204

\bibitem[{{Turtle} {et~al.}(2011){Turtle}, {Perry}, {Hayes}, {Lorenz},
  {Barnes}, {McEwen}, {West}, {Del Genio}, {Barbara}, {Lunine}, {Schaller},
  {Ray}, {Lopes}, \& {Stofan}}]{turtle_eal_2011a}
{Turtle}, E.~P., {Perry}, J.~E., {Hayes}, A.~G., {et~al.} 2011, \science, 331,
  1414

\bibitem[{{Vinatier} {et~al.}(2010){Vinatier}, {B{\'e}zard}, {Nixon},
  {Mamoutkine}, {Carlson}, {Jennings}, {Guandique}, {Teanby}, {Bjoraker},
  {Michael Flasar}, \& {Kunde}}]{vinatier_etal_2010a}
{Vinatier}, S., {B{\'e}zard}, B., {Nixon}, C.~A., {et~al.} 2010, \icarus, 205,
  559

\bibitem[{{Vinatier} {et~al.}(2015){Vinatier}, {B{\'e}zard}, {Lebonnois},
  {Teanby}, {Achterberg}, {Gorius}, {Mamoutkine}, {Guandique}, {Jolly},
  {Jennings}, \& {Flasar}}]{vinatier_etal_2015}
{Vinatier}, S., {B{\'e}zard}, B., {Lebonnois}, S., {et~al.} 2015, \icarus, 250,
  95

\bibitem[{Wood {et~al.}(2010)Wood, Lorenz, Kirk, Lopes, Mitchell, \&
  Stofan}]{Wood2010}
Wood, C.~A., Lorenz, R., Kirk, R., {et~al.} 2010, Icarus, 206, 334

\end{thebibliography}

%\end{linenumbers}

%%%%%%%%%%%%%%%%%%%%%%%%%%%%%%%%%%%%%%%%%%%%%%%%%%%%%%%%%%%%%%%%%%%%%%%%%%%%%%%%%%%%%%%%%%%%%%%%%%%%%%%%%%%%%%%%%%%%%%%%%%%%%%%%%%%%
%% 

\end{document}